\documentclass[12pt]{article}
\usepackage{amsfonts}
\usepackage{amssymb}
\def\hybrid{\topmargin -20pt    \oddsidemargin 0pt
        \headheight 0pt \headsep 0pt
        \textwidth 6.25in       
        \textheight 9.5in       
        \marginparwidth .875in
        \parskip 5pt plus 1pt   \jot = 1.5ex}

\hybrid
\newcommand{\beqn}{\begin{eqnarray}}
\newcommand{\eeqn}{\end{eqnarray}}
\newcommand{\be}{\begin{equation}}
\newcommand{\ee}{\end{equation}}
\newcommand{\non}{\nonumber \\}
\newcommand{\vol}{{\cal V}}
\newcommand{\bi}{\bar{\imath}}
\newcommand{\bj}{\bar{\jmath}}

\newcommand{\pu}{\partial_{\mu}}
\newcommand{\po}{\partial^{\mu}}

\newcommand{\tm}{\tilde{M}}

\newcommand{\tvol}{\tilde{\vol}}
\newcommand{\tg}{\tilde{G}}

\newcommand{\dzzw}{\Phi^{(2)}_{\rm IIA}}
\newcommand{\dzze}{\Phi^{(10)}_{\rm IIA}}
\newcommand{\dhz}{\Phi^{(2)}_{\rm het}}
\newcommand{\dhv}{\Phi^{(4)}_{\rm het}}
\newcommand{\se}{{\scriptsize (E)}}

\def\nN{\hat{N}}
\def\Gi{\hat{G}}
\def\R{({\rm Re}f)}

\def\R{({\rm Re}f)}

\def\rr{r}

\def\Gi{\hat{G}}
\def\nN{\hat{N}}

\begin {document}
\begin{titlepage}
\begin{center}

\hfill hep-th/0011075\\
\vskip 2cm
{\large \bf Type IIA and Heterotic String Vacua in $D=2$}\footnote{Work 
supported by:
DFG -- The German Science Foundation,
GIF -- the German--Israeli
Foundation for Scientific Research,
DAAD -- the German Academic Exchange Service
and the Landesgraduiertenf\"orderung
Sachsen-Anhalt.}

\vskip .5in

{\bf Michael Haack$^a$\footnote{email: {\tt
michael@hera1.physik.uni-halle.de}},
Jan Louis$^a$\footnote{email: {\tt j.louis@physik.uni-halle.de}} 
and Monika Marquart$^{a,b}$\footnote{email: {\tt 
marquart@hera1.physik.uni-halle.de}}} \\

\vskip 0.8cm
{\em $^a$
Fachbereich Physik, Martin-Luther-Universi\"at Halle-Wittenberg,\\
Friedemann-Bach-Platz 6, D-06099 Halle, Germany}

{\em
$^b$  
Department of Physics, Queen Mary and Westfield College\\
 Mile End Road, London E1, U.K.}

\end{center}

\vskip 2.5cm

\begin{center} {\bf ABSTRACT } \end{center}

We study type IIA string theory compactified
on Calabi-Yau fourfolds and heterotic string theory
compactified on  Calabi-Yau threefolds times a two-torus.
We derive the resulting effective theories 
which have two
space-time dimensions and preserve four supercharges.
The duality between such vacua is established at 
the level of the effective theory.
For type IIA vacua with non-trivial Ramond-Ramond
background fluxes a superpotential is generated.
We show that for a specific choice of 
background fluxes and a fourfold
which has the structure of a threefold fibred over a
sphere the superpotential coincides with
the superpotential recently proposed by Taylor and Vafa
in compactifications of type IIB string theory 
on a threefold.

\vskip 1cm

\vfill

November 2000
\vskip 1cm

\end{titlepage}

\section{Introduction} 
Among the various string theory vacua
those with four supercharges in
$D=4$ space-time dimensions 
are phenomenologically the most relevant ones.  
Such vacua have been intensively studied 
and in string perturbation theory
they are constructed for example 
by compactification of the heterotic string 
on a Calabi-Yau threefold $Y_3$ \cite{GSW}.
Some of the non-perturbative
properties of such vacua 
can be obtained from a dual
F-theory compactification on an appropriate 
Calabi-Yau fourfold $Y_4$ \cite{V}. 

Sometimes the non-perturbative 
features of a given string vacuum
simplify in lower dimensions
as a consequence of a simpler 
dual description. 
For example the dual of 
the heterotic string compactified on 
$Y_3\times S^1$ is given by M-theory
compactified  on $Y_4$ while the dual of the 
heterotic string compactified on
$Y_3\times T^2$ is type IIA compactified 
on $Y_4$ \cite{V}.
 
In this paper we focus on string vacua in $D=2$
with four supercharges,
the heterotic string compactified on
$Y_3\times T^2$ and the dual 
type IIA compactified on $Y_4$.
In the first part of the paper
(section 2) we derive the low energy effective
theories and establish the duality at the level
of the effective theory. We display
an explicit map between the heterotic
variables and their dual type IIA counterparts.
In the second part of the paper (section 3)
we consider
type IIA backgrounds where non-trivial 
RR-fluxes have been turned on. 
The specific case of a 4-form flux  
can be used to construct a consistent
compactification if the Euler number 
of the fourfold is non-vanishing \cite{SVW}.
The presence of such fluxes leads to 
a non-trivial superpotential $W$  and a 
supersymmetric vacuum is found only on a subspace
of the moduli space \cite{M,L,K,GVW,DRS,G,GSS}.

The superpotential receives two distinct
contributions $W$ and $\tilde W$ \cite{GVW,G}. 
$\tilde W$ arises
from wrapping  D$p$-branes on holomorphic
$p$-cycles and depends on the deformations
of the (complexified) K\"ahler class,
i.e.\ the harmonic $(1,1)$-forms on $Y_4$.
$W$ on the other hand is generated by
 wrapped $D4$-branes on special
Lagrangian cycles and thus depends on
the deformations of the complex structure
or the harmonic $(1,3)$-forms on $Y_4$.
$\tilde W$ receives 
quantum corrections on the worldsheet whereas
$W$ is determined by a classical computation.
Mirror symmetry relates the two
superpotentials and demands
$\tilde W(Y_4) = W(\tilde Y_4)$,
where $\tilde Y_4$ is the mirror fourfold of $Y_4$.

In section 3.2 we show that 
for a particular class of
fourfolds which have the structure of 
a Calabi-Yau threefold $Y_3$ fibred over
a base $\mathbb{P}_1$
the superpotential is determined 
by the prepotential ${\cal F}$
of $Y_3$ if the background fluxes are
suitably chosen and the $\mathbb{P}_1$ is 
taken to be large. 
Curiously this superpotential coincides
with the superpotential derived in ref.\ \cite{TV}.
It arises in type II compactification on a Calabi-Yau 
threefold with non-vanishing fluxes 
studied in refs.\ \cite{PS,JM,TV,M2}. 
It also is very closely related to the BPS-mass formula
studied in refs.\ \cite{CDFP,CLM} and the entropy 
of $N=2$ extremal black holes studied in
refs.\ \cite{FK,TM}.

In section 3.3 we consider fourfolds which have the 
structure of a $K3$ manifold fibred over a 
(complex) two-dimensional Hirzebruch surface $\mathbf F_n$.
In the limit of a large base  $\mathbf F_n$
the superpotential is determined by 
the prepotential of $K3$ fibred threefolds
in the large base limit.
In this case the duality to heterotic vacua
discussed in section 2.3 can be used
to also determine the heterotic superpotential.

Finally, some of the technical details are 
deferred to three appendices.

\section{Heterotic -- type IIA duality in $D=2$ without 
superpotential}

\subsection{Heterotic vacua in $D=2$}
In this section we perform a Kaluza-Klein
reduction of a generic $D=4,N=1$  heterotic
effective theory on a torus $T^2$.
This results in an effective theory in $D=2$
with $(2,2)$ supersymmetry. 
For the purpose of this paper it is sufficient
to focus only on the vector multiplets and the
 chiral moduli multiplets in the $D=4$ effective 
action and 
ignore all charged matter multiplets.
The bosonic part of the effective Lagrangian in this case reads
\be
{\cal{L}}_{\rm het}^{(4)} = \frac{1}{2} R^{(4)} - 
G_{I \bar J}^{(4)} (\Phi,\bar{\Phi}) \partial_m
\Phi^I \partial^m {\bar \Phi}^{\bar J} 
- \frac{1}{4} \mbox{Re}  f_{ab}(\Phi) 
F^{a}_{mn} F^{b mn} 
+ 
\frac{1}{4} \mbox{Im} f_{ab}(\Phi) F^{a}_{mn} 
\tilde F^{b mn} ,
\label{het4d}
\ee
where  $m,n=0,\ldots,3$, 
$\Phi^I$ are the moduli fields
and $F^{a}_{mn}$ 
is the field strength of the gauge bosons
$A_m^a, a=1,\ldots, {\rm dim}(G)$. 
The $f_{a b}(\Phi)$ are holomorphic functions
of the moduli and
the metric $G_{I \bar{J}}^{(4)}$ is a K\"ahler metric with 
K\"ahler potential $K^{(4)}(\Phi,\bar{\Phi})$. 
At the string tree level $K^{(4)}$ and  $f_{a b}$
are further constrained
 to obey\footnote{For 
the general case see appendix \ref{hetred}.} 
\be \label{tree}
K^{(4)}_{\rm het}(\phi,\bar{\phi},S,\bar S) 
=\tilde K^{(4)}_{\rm het}(\phi,\bar{\phi})
-\ln[i(\bar S- S)]\
,
\qquad
f_{ab}=-i S\delta_{ab}\ ,
\ee
where $S$ is the heterotic dilaton 
and the $\phi$ denote all moduli 
except the dilaton.\footnote{Contrary to the standard
heterotic convention we choose to identify the dilaton
with $\mbox{Im} S$ in order to simplify the 
discussion in section~3.}

The next step is to perform a Kaluza-Klein 
reduction on a torus. 
The chiral multiplets
survive the reduction unaltered and continue
to contain a complex scalar field as bosonic
component. The vectors on
the other hand decompose according to 
\be \label{ahat}
A^a_n d x^n = A^a d \zeta 
+ \bar A^a d \bar \zeta + A^a_\mu dx^\mu\ ,
\ee 
where $\mu=0,1$ and $\zeta$ is 
the complex coordinate on the 
torus. The vectors $A_\mu^a$ 
do not have any dynamical degrees of freedom 
and play the role of 
auxiliary fields. Thus the physical bosonic components of 
the
vector multiplets in $D=2$ consist of 
the complex scalars $A^a$.
These scalars transform in the adjoint
representation of the gauge group $G$
and thus their vacuum expectation values break
$G$ to its maximal Abelian subgroup 
$U(1)^{{\rm rank}(G)}$. 
With a slight abuse of notation
the index $a$ from now on only takes the values
$a=1,\ldots,{\rm rank}(G)$. 
On this Coulomb branch each vector 
occurs in the Lagrangian 
only via its Abelian field strength. In this case 
the field strength can be replaced 
by an auxiliary scalar field \cite{GGW2} and 
one ends up 
with the field content of a twisted chiral multiplet
\cite{GHR}. 
This is one of the two 
different kinds of matter multiplets 
which can occur in $D=2$ theories 
with $(2,2)$ supersymmetry, the other one 
being the chiral multiplet. 
We later use the fact that under certain conditions they 
are  dual to each other. 
More precisely, if the Lagrangian is invariant under a 
Peccei-Quinn shift symmetry of a chiral 
multiplet one can perform a duality 
transformation replacing a chiral 
by a twisted chiral multiplet 
- and vice versa  \cite{GHR}.

Finally, we need to decompose the four-dimensional
space-time metric. Instead of using the Einstein
metric implicit in eq.\ (\ref{het4d}) it turns out to
be more convenient to work in the string frame 
metric and decompose it according to \footnote{The reason 
for taking the string-frame metric is that the variables which we get are
exactly those we need in section \ref{hetiia} in order to establish the 
map to the type IIA variables in which the superpotential is naturally
expressed. This will become clear in the following sections.} 
\be \label{gmn}
g_{mn}dx^mdx^n = g^{(2)}_{\mu\nu}dx^{\mu}dx^{\nu}+
b_{\mu i} dx^{\mu}dx^{i}+
h_{ij}dx^{i}dx^{j}\ , \quad i,j = 2,3 \ ,
\ee
where $g^{(2)}_{\mu\nu}$ and $b_{\mu i}$ have no
physical degrees of freedom.
The real coordinates $x^2,x^3$ are related to
the complex $\zeta$ of eq.\ (\ref{ahat}) via
$d \zeta = dx^2 + \tau dx^3$
where $\tau$  is the complex structure 
modulus of the torus
\be \label{tau}
\tau \equiv  \left( \frac{h_{12}}{h_{11}} 
+ i\, \frac{\sqrt{h}}{h_{11}} \right),
\ee
and $h$ is the determinant of  $h_{i j}$.

We omit the details of 
the standard Kaluza-Klein compactification 
on the torus and only present
the resulting two-dimensional effective
Lagrangian
\beqn \label{het2}
{\cal L}_{\rm het}^{(2)} & = & e^{-2 \dhz} 
\left[ -\frac{1}{2} \pu \po \sigma 
+ \frac{\pu \tau \po \bar \tau}{(\tau - \bar \tau)^2}
- \frac{1}{4} h^{-1} \pu \sqrt{h} \po \sqrt{h} 
- \tilde G_{I \bar J}^{(4)} \pu \phi^I \po \bar \phi^{\bar J} 
\right. \\
& -&\!\! \left.  i (\tau - \bar \tau)^{-1} \sqrt{h}^{-1} 
D_\mu \bar n^a D^\mu n^a 
- h^{-1} \Big(\partial_\mu P + \frac{1}{2} 
(\tau - \bar \tau)^{-1}[\bar n^a 
D_\mu n^a - n^a D_\mu \bar n^a] \Big)^2 
\right]\ , \nonumber
\eeqn
where we defined
\be \label{dhz}
e^{-2 \dhz} = \frac{\sqrt{h}}{2 i}\, (S - \bar S) \ ,
\qquad
n^a = -i (\tau - \bar \tau) \bar A^a\ .
\ee
The covariant derivatives of $n^a$ are given by
\be \label{dmuna}
D_\mu n^a \equiv \pu n^a - 
(\tau - \bar \tau)^{-1} \pu \tau \big(
n^a - \bar n^a \big) \ ,
\ee
and the scalar field $P$ in the Lagrangian 
(\ref{het2}) is the 
dual of the axion ${\rm Re}S$.\footnote{ 
More precisely we added the term
$\partial_\mu P \epsilon^{\mu \nu} \partial_\nu {\rm Re}S $ 
to the reduced action and eliminated 
{\rm Re}S by its equation of motion.
One can also verify that $2P$
is the $B_{23}$ component of the 
$D=4$ antisymmetric tensor in the 
compactified directions.}
Moreover we have chosen the conformal gauge for 
the $D=2$ metric
\be
g^{(2)}_{\mu \nu} = e^\sigma e^{2\dhz} \eta_{\mu \nu}\ .
\ee
Finally, the index $I$ now denotes all moduli except 
the dilaton and  $\tilde G_{I \bar J}^{(4)}$ is the 
K\"ahler metric of the K\"ahler potential 
$\tilde K^{(4)}_{\rm het}$ defined in eq.\ (\ref{tree}).

Defining the complexified K\"ahler 
modulus as 
\be \label{rho}
\rho \equiv i \sqrt{h} + (\tau - \bar \tau)^{-1}
n^a (n^a - \bar n^a) + 2 P
\ee 
one verifies that the Lagrangian (\ref{het2})
can be written in the form 
\be 
{\cal L}_{\rm het}^{(2)} =  e^{-2 \dhz} 
\left[ -\frac{1}{2} \pu \po \sigma - G_{\bar \Lambda 
\Sigma} \pu \bar Z^{\bar \Lambda} \po Z^\Sigma 
\right]\ ,
\label{l22}
\ee 
where the $Z^\Sigma$ 
denote  $Z^\Sigma = (\phi^I, \tau, \rho, n^a)$.
The $Z^\Sigma$ are the proper K\"ahler 
coordinates in that in these coordinates
$G_{\bar \Lambda \Sigma} = 
\bar\partial_{\bar\Lambda}\partial_{\Sigma}K^{(2)}_{{\rm het}}$
with 
\be\label{k2het}
K^{(2)}_{{\rm het}} = \tilde{K}^{(4)}_{{\rm het}}
(\phi, \bar \phi) 
- \ln[(n^a - \bar n^a)^2 
- (\rho - \bar \rho)(\tau - \bar \tau)].
\ee

The form (\ref{l22}) of the Lagrangian
explicitly shows that both $\dhz$
and $\sigma$ are unphysical degrees of freedom
\cite{DNRT} while $\rho$ and $\tau$ are 
propagating degrees of freedom.
The four physical degrees of freedom 
in $\rho$ and $\tau$ are 
related to the four-dimensional graviton and the dilaton
$S$.

In this parameterization the modular
group of the torus $SL(2,\mathbb Z)\times
SL(2,\mathbb Z)$ acting on $\tau$ and $\rho$
as fractional linear transformation is
manifest. The first $SL(2,\mathbb Z)$ acts as 
\beqn \label{sl2z}
\tau\to\frac{a\tau+b}{c\tau+d}, \qquad \rho\to \rho - 
\frac{c\, n^a n^a}{c\tau+d},\qquad n^a \to \frac{n^a}{c\tau+d}, \qquad
 \left( \begin{array}{cc}
                     a & b \\
                     c& d
                     \end{array} \right)\in SL(2,\mathbb Z) 
\eeqn
and the action of the second $SL(2,\mathbb Z)$ is obtained by
exchanging $\tau$ and $\rho$.

$\tau$ and $\rho$ both reside in vector
or twisted chiral multiplets since in the reduction
procedure they come along
with the Kaluza-Klein vectors.\footnote{The modulus 
$\rho$ is the dual of the chiral 
field $S$ and therefore twisted chiral. A further argument 
that $\tau$ is indeed part of a twisted chiral multiplet 
is given in appendix \ref{hetred}.} 
{}From eq.\ (\ref{k2het}) we learn that  
the moduli space factorizes into chiral multiplets $\phi^I$ and twisted chiral
multiplets $(\tau,\rho,n^a)$ which 
span the coset space $SO(2,2+r)/SO(2)\times SO(2+r)$
where $r=\mbox{rank}(G)$.
In this case the two-dimensional
$(2,2)$ supersymmetric $\sigma$-model
is known to be K\"ahler \cite{GHR}, where the K\"ahler potential is a sum of
two terms, one depending on the chiral multiplets and one depending on the
twisted chiral multiplets.
However, in more general situations
the metric  is not K\"ahler 
but nevertheless 
the Lagrangian can be expressed in terms of  
real 
functions similar to the K\"ahler potential. 
This situation occurs if one goes beyond the tree level
approximation and includes 
string loop corrections.
Some results about the general case are presented 
in appendix \ref{hetred}.
For completeness this appendix also contains a different derivation of the
$D=2$ effective theory by compactifying the 
$D=3$ Lagrangian given in ref.\ \cite{HL} on a circle.


\subsection{Type IIA theory on Calabi-Yau fourfolds}
In this section we discuss the Kaluza-Klein reduction
of type IIA on Calabi-Yau fourfolds.
The
starting point is the low energy effective action in $D=10$ 
in the string-frame 
\beqn
S_{IIA}^{(10)} & = & \int d^{10}x \, 
\sqrt{-g^{(10)}} e^{-2 \dzze} \left(\frac{1}{2} R^{(10)} 
+ 2 \partial_{M} \dzze \partial^{M} \dzze 
- \frac{1}{4} |H_{3}|^{2} \right) \non
 & &  - \frac{1}{4} \int d^{10}x \, 
\sqrt{-g^{(10)}} \left(|F_{2}|^{2} + |\tilde{F}_{4}|^{2} \right) 
 - \frac{1}{4} \int B_{2} \wedge F_{4} 
\wedge F_{4}\ ,
\label{IIA10d}
\eeqn
where $\dzze$ is the ten-dimensional dilaton,
$H_{3}  =  d B_{2}$ is the field strength
of the antisymmetric tensor $B_2$ and
$F_{2} =  d C_{1}$ the field strength of the 
RR vector $C_1$. 
$F_{4} =  d C_{3}$ is the field strength
of the RR 3-form $C_3$ and we 
use the abbreviation 
$\tilde{F}_{4} =  F_{4} - C_{1} \wedge H_{3}$.
For further conventions on the notation see appendix \ref{notation}.

In eq.\ (\ref{IIA10d}) only
the leading terms
of $S_{IIA}^{(10)}$  are displayed and higher derivative
couplings are suppressed. 
In particular the term proportional to
$\int B_2 \wedge X_8, \
X_8 \sim   4 \mbox{tr} R^4 - (\mbox{tr} R^2)^2 $
imposes a consistency condition on the
compactification \cite{SVW}. 
The absence of a $B_2$-tadpole requires
\be\label{ccond}
-\int_{Y_4} X_8\ =\ {\chi\over 24}\ =\ 
n+ \frac1{8\pi^2} \int_{Y_4} F_4\wedge F_4\ ,
\ee
where $n$ is the number of space-time filling strings.
In this section we focus on the case $\chi=n=F_4=0$
and return to the case of non-trivial $F_4$
in section~3.

The spectrum of the $D=2$ theory is determined by the deformations 
of the Calabi-Yau metric and the expansion of $C_1,B_2$ and $C_3$ 
in terms of the non-trivial forms of $Y_4$. 
The deformations of the metric comprise $h^{1,1}$
real K\"ahler deformations $M^A, A=1,\ldots, h^{1,1}$
and $h^{1,3}$ complex deformations
$Z^\alpha, \alpha=1,\ldots, h^{1,3}$ of the complex
structure.
Since vectors contain no physical degree
of freedom in $D=2$ and since there are no 
1-forms on $Y_4$ the 1-form
$C_1$ does not contribute any massless
mode in $D=2$. $B_2$ leads to
$h^{1,1}$ real scalar fields $a^A$ while
$C_3$ contributes $h^{1,2}$ complex scalars
$N^I, I = 1,\ldots, h^{1,2}$.
The (1,1)-moduli reside in twisted chiral
multiplets\footnote{Strictly speaking these
multiplets are vector multiplets containing 
in addition the vectors arising from expanding the 
3-form in terms of the (1,1)-forms of $Y_4$. But 
as noted in the last section, 
a vector multiplet is related to a twisted chiral
multiplet, differing only in the auxiliary field 
content.}  
while all other scalars 
are members of chiral multiplets.

For simplicity let us discuss here only 
the case 
where the $(2,1)$-moduli are frozen to some fixed 
value and postpone the discussion
of the general case to appendix \ref{IIAred}.
Compactification of the action (\ref{IIA10d})
results in
\be
\label{2a2}
{\cal L}_{IIA}^{(2)} =   
\sqrt{-g^{(2)}}\, e^{-2 \dzzw} \left(\frac{1}{2} R^{(2)} 
+ 2 \partial_{\mu} \dzzw \partial^{\mu} \dzzw 
-G_{A \bar B} \partial_{\mu} t^{A} 
\partial^{\mu} \bar{t}^{\bar B} - G_{\bar{\alpha} \beta} \partial_{\mu} 
\bar{Z}^{\bar{\alpha}} \partial^{\mu} Z^{\beta} \right)\ ,
\ee
where the following definitions have been used
\beqn
e^{-2 \dzzw} & \equiv &  e^{-2 \dzze} \vol\ , \non
t^A & \equiv & \frac{1}{\sqrt{2}} (a^A +i M^A)\ , \non
G_{A \bar B} & \equiv & \frac{1}{2 {\vol}} 
\int_{Y_4} e_{A} \wedge \star e_{B} = 
-\partial_A \bar\partial_{{\bar B}} \ln \vol \ , \non
G_{\alpha \bar{\beta}} & \equiv & 
- \frac{\int_{Y_4}
 \Phi_{\alpha} \wedge \bar{\Phi}_{\bar{\beta}}}
{\int_{Y_4} \Omega \wedge 
\bar{\Omega}} = -\partial_{\alpha} 
\bar\partial_{{\bar \beta}} 
\ln \Big[ \int_{Y_4} \Omega \wedge 
\bar{\Omega} \Big]\ . \label{gab}
\eeqn
$e_A$ denotes a basis 
for the (1,1)-forms of $Y_4$, $\Phi_\alpha$ 
 a basis for its
(3,1)-forms and $\Omega$ is its unique (4,0)-form.
$\vol$ is the volume of the fourfold
which can be expressed in terms of the K\"ahler form
$J=M^A e_A$ as
\be\label{vol}
\vol = \frac{1}{4!} \int_{Y_4} J\wedge J\wedge J\wedge J
=
\frac{1}{4 \cdot 4!}\, d_{ABCD}(t^A - \bar t^A)
(t^B - \bar t^B)(t^C - \bar t^C)(t^D - \bar t^D)\ ,
\ee 
where  $d_{ABCD}$ are  the classical intersection numbers 
$d_{ABCD} = \int_{\footnotesize Y_4} e_A 
\wedge e_B \wedge e_C \wedge e_D$ .

As in the heterotic case the moduli space 
factorizes into chiral and twisted chiral multiplets, i.e. 
it is K\"ahler although both kinds of multiplets 
occur.
The K\"ahler potential can be read off from (\ref{gab}) 
and is given by
\be\label{k22a}
K^{(2)}_{IIA} =  - \ln \big( \int_{Y_4} \Omega \wedge 
\bar{\Omega} \big) - \ln \vol\ . 
\ee
If (2,1)-moduli 
arising from $C_3$ are taken into account
the metric ceases to be K\"ahler but
still can be expressed in terms of a real 
function. This situation is discussed 
in appendix~\ref{IIAred}.\footnote{ 
The same appendix also presents a different
derivation of the $D=2$ effective theory
where an $S^1$ reduction of the three-dimensional
effective theory  of ref.\ \cite{HL}
obtained as 
compactification of $D=11$ 
supergravity on Calabi-Yau fourfolds $Y_4$
is performed.
In this case all scalar fields appear naturally
as members of chiral multiplets and as a 
consequence the metric in these coordinates
is always K\"ahler. The K\"ahler potential
of the $D=2$ effective theory
coincides with the K\"ahler potential
of the $D=3$ effective theory.}

Mirror symmetry is believed to be a property
of all Calabi-Yau manifolds and for fourfolds
it exchanges $(1,1)$- and $(1,3)$-forms \cite{GMP,M,KLRY}.
As it stands the K\"ahler potential (\ref{k22a}) does not
share this symmetry due to the fact
that the derivation used the supergravity
approximation which is only valid if the size of
the manifold is large compared to the string length
(i.e.\ in the large radius approximation). 
For `small' fourfolds
worldsheet instantons correct the K\"ahler potential
of the $(1,1)$-moduli and are believed to 
render (\ref{k22a})
mirror symmetric.


\subsection{Heterotic -- type IIA duality in D=2}
\label{hetiia}

In order to establish the duality relationship between 
the heterotic and type IIA variables we specify 
the compactification manifolds.
The Calabi-Yau fourfold is taken to be a 
$K3$-fibration over 
a large Hirzebruch surface $\mathbb{F}_n$ 
with $n$ even while the Calabi-Yau threefold 
on the heterotic side is an elliptic fibration 
over the same base. 
In the large base limit  
the type IIA K\"ahler potential simplifies as \cite{HL}
\be
K^{(2)}_{IIA} = -\ln\vol \ \to\ 
-\ln\big[(t^U - \bar t^U)(t^V - \bar t^V) 
\eta_{{\hat \imath}{\hat \jmath}}
(t^{\hat \imath} - \bar t^{\hat \imath})
(t^{\hat \jmath} - \bar t^{\hat \jmath})\big]
\ ,
\label{kl}
\ee
where $t^U$ and $t^V$ are the moduli of
the base while the $t^{\hat \imath}$
denote the moduli of the $K3$-fibre
except those from reducible bad fibres.
$\eta$ is the intersection matrix of $K3$
\be\label{etadef}
\eta_{{\hat \imath}{\hat \jmath}} = \left( \begin{array}{ccc}
                     
                      0 & 1/2 & \\
                      1/2 & 0 & \\
                        &   & - {\bf I}
                     \end{array} \right).
\ee

On the heterotic side 
 the K\"ahler potential (\ref{k2het}) becomes
in the large base limit
\be \label{k2lb}
K^{(2)}_{{\rm het}} = -\ln\big[(U - \bar U) (V - \bar V) 
((\tau - \bar \tau)(\rho - \bar \rho) 
- (n^a - \bar n^a)^2) \big]\ ,
\ee
where $U$ and $V$ are the two base moduli.
Comparing (\ref{kl}) and (\ref{k2lb})
it is tempting to equate the two expressions.
However, this would map the twisted chiral 
superfields $t^U$ and $t^V$ to chiral superfields
$U$ and $V$. Hence one 
has to first perform an additional 
duality transformation on the base moduli
$U,V$ so that all heterotic variables are
twisted chiral like their type IIA counterparts. 
This is possible in the large base limit
by defining
$c^\nu_U = -\epsilon^{\mu \nu} \pu {\rm Re} U$ and
$c^\nu_V = -\epsilon^{\mu \nu} \pu {\rm Re}V$ and 
add the Lagrange multipliers $P_U,P_V$ via
$- c^\mu_U \pu P_U - c^\mu_V \pu P_V$ 
to the action. Using the 
equations of motion to eliminate $c^\mu_U$ and 
$c^\mu_V$, defining the coordinates
\be
u \equiv 2 P_U + i\, e^{-2 \dhz} ({\rm Im} U)^{-1} 
\ , \qquad 
v \equiv 2 P_V + i\, e^{-2 \dhz} ({\rm Im} V)^{-1} 
\ee
and
redefining the $D=2$ metric
\be
g^{(2)}_{\mu \nu} = e^\sigma e^{4 \dhz} {\rm Im} U {\rm Im} V 
\eta_{\mu \nu}
\ee
one derives the following form of the 
K\"ahler potential 
\be
\hat{K}^{(2)}_{{\rm het}} = 
-\ln[(u - \bar u)(v - \bar v)
((\tau - \bar \tau)(\rho - \bar \rho) 
- (n^a - \bar n^a)^2)].
\label{kx}
\ee
In this form all the coordinates belong to 
twisted chiral multiplets.  
Hence the duality map 
relates\footnote{The fact that it is really 
$u$ and $v$ which should be mapped to $t^U$ 
and $t^V$ can heuristically be understood from the
duality in $D=6$ fibred over a four-dimensional base 
manifold $B$. It yields $\vol_B^{\rm IIA} = e^{-4 \Phi^{(6)}_{\rm het}} 
\vol_B^{\rm het} = e^{-2 \Phi^{(6)}_{\rm het}} e^{-2 \dhz}$ and 
similarly $\vol_B^{\rm het} = e^{-2 \Phi^{(6)}_{\rm IIA}} 
e^{-2 \dzzw} = e^{2 \Phi^{(6)}_{\rm het}} 
e^{-2 \dzzw}$, where all volumes are those 
of the base measured in the corresponding string-frame metric.
Using (\ref{deqd}) one verifies immediately 
$\vol_B^{\rm IIA}=(\vol_B^{\rm het} e^{4 \dhz})^{-1}$.}
\beqn
\left\{ t^{\hat \imath} \right\} & \leftrightarrow & \left\{ 
\tau, \rho, n^a \right\} , \non
\left\{ t^U, t^V \right\} & \leftrightarrow & 
\left\{ u, v \right\}.
\label{abb}
\eeqn

As we discussed in section 2.1 
the heterotic theory has an
 $SL(2,\mathbb Z)\times SL(2,\mathbb Z)$
symmetry (\ref{sl2z}) which is just the modular 
symmetry of the torus. From the map (\ref{abb}) 
one learns that in the type IIA theory this symmetry
has to be a property of $K3$-fibred
fourfolds in the large base limit.
This is precisely the same situation one encounters
in the four-dimensional duality relating
type IIA compactified on $K3$-fibred threefolds 
to heterotic vacua compactified on $K3\times T^2$
\cite{D4rev}.

A similar analysis was performed in $D=3$ in ref.\
\cite{HL}. 
As mentioned in the previous sections one can 
derive the two-dimensional effective actions via 
the detour over $D=3$. In this case the scalars are 
members of chiral multiplets only and therefore 
parameterize a K\"ahler manifold with the 
same K\"ahler potential as in $D=3$. Thus the map 
derived in \cite{HL} in the corresponding coordinates 
continues to hold in $D=2$. However, for the 
purpose of this paper the map (\ref{abb}) 
turns out to be more useful.
For further details we refer the reader to appendix C.
\section{Type IIA string with background fluxes}
\subsection{The superpotential in IIA}\label{IIA}
In this section we consider compactifications of type IIA string
theory on Calabi-Yau fourfolds in a background where non-trivial
RR-fluxes have been turned on and simultaneously all $(1,2)$ moduli have been
frozen. In refs.\ \cite{GVW,G} it was shown 
that the low energy effective theory can be described
by a $(2,2)$-supersymmetric Lagrangian with a non-trivial potential
which depends on the background fluxes. 
The potential $V$ is derived from two superpotentials $W$ and 
$\tilde W$ \cite{GGW}
\be
V \sim e^{K^{(2)}_{IIA}} \left(G^{-1 A\bar B}
D_{A}\tilde WD_{\bar{B}}\bar{\tilde W}
+G^{-1\alpha\bar\beta}D_{\alpha}{W}
D_{\bar\beta}\bar{{W}}-|W|^2-|\tilde{W}|^2\right)\ ,
\ee
where $W$ depends on the chiral complex structure scalars $Z^\alpha$ 
and $\tilde W$ on the twisted chiral K\"ahler scalars $t^A$. 
The K\"ahler covariant derivatives are defined as
\be
D_{A}\tilde W =\partial_A \tilde W + \tilde W 
\partial_A K^{(2)}_{IIA} \ ,\qquad
D_{\alpha}{W} = \partial_\alpha  W +  W \partial_\alpha K^{(2)}_{IIA}\ .
\ee

Since D-branes are the magnetic sources of RR-fluxes 
the generation of the superpotentials can also be understood from
wrapping the D-branes in the IIA theory on supersymmetric cycles.
More specifically, wrapping a D4-brane on a 
four-dimensional special Lagrangian cycle
generates the superpotential \cite{GVW,G}\footnote{As 
argued in \cite{GVW} the change of the superpotential 
when crossing the brane in the two-dimensional space-time
is equal to the volume of the four-cycle $C^{(4)} \in 
H_4(Y_4, \mathbb{Z})$ wrapped by the D4-brane: $\Delta W =
\int_{C^{(4)}}\Omega=\frac{1}{2\pi}\int_{Y_4}\Omega\wedge\Delta F_4$. 
The change of the 4-form flux $\frac{\Delta F_4}{2\pi}
\in H^4(Y_4,\mathbb{Z})$ is  Poincar\'e 
dual to the four-cycle $C^{(4)}$. 
Note that $\frac{F_4}{2\pi}$ itself does in general not take values 
in $H^4(Y_4,\mathbb{Z})$ but $\frac{F_4}{2\pi}-\frac{p_1}{4}$, where $p_1$
is the first Pontryagin class \cite{W1}.}
\be\label{W}
W=\frac{1}{2\pi}\int_{Y_4}\Omega\wedge F_4\ ,
\ee 
where $F_4$ denotes the RR 4-form flux.

A second superpotential is generated by wrapping
D$p$-branes, $p=0,2,4,6,8$,  on holomorphic cycles 
$C^{(p)}\in H_{p}(Y_4, \mathbb{Z})$ with the same real dimension
$p$ \cite{GVW,G}
\be\label{Wtilde}
\tilde W_{\rm cl}=\frac{1}{2\pi} \int_{Y_4}\left(t\wedge t \wedge t \wedge t\ F_0+t\wedge
  t\wedge t\wedge F_2+ t\wedge t\wedge F_4+t\wedge F_6+F_8\right)\ , 
\ee
where $t= t^A e_A$ and
$\frac{F_p}{2\pi}\in H^p(Y_4, \mathbb{Z})$ 
is the RR $p$-form flux  which is Poincar\'e dual
to the $(8-p)$-cycle $C^{(8-p)}\in H_{8-p}(Y_4,\mathbb{Z})$
(for $p=4$ see the last footnote).\footnote{It should
be possible to derive $W,\tilde W$ also from
a KK-reduction as outlined in the previous section
where appropriate fluxes have been turned on 
\cite{GSS,HL2}.}

The effective theory has $(2,2)$ supersymmetry
in a Minkowskian background if
\be\label{susycond}
D_{A}\tilde W|_{\rm min}=D_{\alpha}{W}|_{\rm min}
=\tilde{W}|_{\rm min}=W|_{\rm min}=0
\ee
holds. Depending on the background fluxes
this puts a severe constraint on the moduli space and
for some fluxes no supersymmetric vacuum
exists at all. 
For example, for the $W$ of eq.\ (\ref{W})
the supersymmetry condition (\ref{susycond})
implies \cite{BB} 
\be\label{f4}
 F_4^{(0,4)}= 0 = F_4^{(1,3)}\ ,
\ee
where the last equation arises from the fact
that $\partial_\alpha\Omega$ takes
values in $H^{4,0}$ and in $H^{3,1}$. 
Since the Hodge decomposition of
$H^4$ depends on the complex structure,
eq.\ (\ref{f4}) is a strong constraint on the
moduli space of the complex structure.
It leaves only the subspace 
of complex structure deformations 
which respect (\ref{f4}) as the physical
moduli space.

The superpotential
$\tilde W$ receives quantum corrections on the 
worldsheet while $W$ is exact.
Mirror symmetry demands that once all quantum corrections
are properly taken
into account the two superpotentials should 
obey \cite{OOY,L,G} 
\be\label{weqw}
\tilde W(Y_4)= \tilde W_{\rm cl}(Y_4)
+\mbox{quantum corrections}=
W(\tilde Y_4)\ ,
\ee
where $\tilde Y_4$ is the mirror fourfold
of $Y_4$. The quantum corrections can be derived 
by computing $W$ on the mirror manifold and performing 
the mirror map. 
This is of interest since in the dual 
heterotic vacuum the $(1,1)$-scalars correspond
to the heterotic variables $\{\tau,\rho,n^a\}$
(c.f.\ (\ref{abb}))
which are related to the four-dimensional heterotic
dilaton. 
Thus the quantum corrected $\tilde W(Y_4)$ possibly
encodes non-trivial information about space-time 
quantum corrections of four-dimensional heterotic
vacua. 
As we will see 
the superpotential evaluated
in the large base limit 
of  $K3$-fibred 
fourfolds which are simultaneously $Y_3$-fibred
is determined 
by the prepotential ${\cal F}$
of $Y_3$
if the background values for $F_p$ are 
suitably chosen.

\subsection{The superpotential for threefold-fibred fourfolds} 

Let us first introduce the notion of    
the vertical primary subspace \cite{GMP,M}.
This is the
subspace of $\oplus_{k=0}^{d} H^{(k,k)}(Y_d)$ obtained by taking all 
wedge products
of the $(1,1)$-forms.
The horizontal primary subspace of 
$\oplus_{k=0}^{d} H^{(d-k,k)}(Y_d)$
can be obtained from the vertical primary subspace 
of the mirror manifold via mirror 
symmetry
\cite{GMP,M,KLRY}.\footnote{The horizontal primary 
subspace is generated by successive derivatives of the 
holomorphic $(d,0)$-form $\Omega$ \cite{AS,GMP}.} 
Let us consider the class of 
fourfolds which are $Y_3$-fibred  over
a base $\mathbb{P}^1$ \cite{M}. Assuming the absence of reducible
bad fibres in this $Y_3$-fibration
the vertical primary subspace can be obtained by 
taking the wedge product of the elements of the vertical primary 
subspace of $Y_3$ with the zero- or $(1,1)$-forms
on the $\mathbb P^1$ base. This leads to the following 
Hodge numbers of
the vertical primary subspace of $Y_4$
\beqn \label{hnu}
h^{(0,0)}(Y_4)&=&1\ =\ h^{(4,4)}(Y_4) , \nonumber \\
h^{(1,1)}(Y_4)&=&h^{(1,1)}(Y_3)+1\ =\ 
h^{(3,3)}(Y_4)\ , \\
h^{(2,2)}_V(Y_4)&=&2h^{(1,1)}(Y_3)\ . \nonumber
\eeqn
Except for $h^{(2,2)}_V$ the Hodge numbers of the 
vertical primary subspace coincide with those of the full
vertical cohomology. In the following we also use the 
formulation of the vertical primary  subspace in terms of 
the dual homology. In the homology the vertical primary 
 subspace is obtained by joining the even-dimensional 
cycles of $Y_3$ with the zero- or two-cycles 
of the base $\mathbb P^1$.

Mirror symmetry implies the following relations
for the Hodge numbers of the horizontal primary subspace
of $\tilde Y_4$ 
\beqn \label{mirrhnu}
& &h^{(0,4)}(\tilde Y_4)=h^{(4,4)}(Y_4), \ \ \ h^{(4,0)}(\tilde Y_4)=h^{(0,0)}(Y_4), \\
& &h^{(3,1)}(\tilde Y_4)=h^{(1,1)}(Y_4), \ \ \ h^{(2,2)}_H(\tilde Y_4)=h^{(2,2)}_V(Y_4). \nonumber
\eeqn
The members of the vertical respectively the
horizontal primary cohomology are observables
in the A- respectively B-model. The A- and the B-model are  two topological
sigma-models with the Calabi-Yau manifold $Y_4$ as a target space,
which are obtained by twisting the worldsheet sigma-model in two different
ways \cite{W3}. The observables and correlation functions of the A-model 
on $Y_4$ are related via mirror symmetry to those of the 
B-model on $\tilde Y_4$ and vice versa.

\subsubsection{Computation of $W(\tilde Y_4)$ and 
$\tilde W(Y_4)$}
\label{Wsec}
Our goal is to compute 
$\tilde W(Y_4)$ in the large base limit using 
mirror symmetry and eqs.\ (\ref{weqw}) and (\ref{W}). 
This is possible if we choose
the background fluxes $F_p$ on $Y_4$ 
to lie in the vertical primary subspace.
These fluxes are mapped to elements of the horizontal
primary subspace on the mirror manifold ${\tilde Y}_4$.
The dimension  $h_H^4(\tilde Y_4)$ of the horizontal
primary subspace is given by
\be
h_H^4=2h^{(4,0)}(\tilde Y_4) + 2 h^{(3,1)}(\tilde Y_4) +
h^{(2,2)}_H(\tilde Y_4) = 4 (h^{(1,1)}(Y_3)+1)\ .
\ee
Let us denote a basis for the  horizontal
primary homology of ${\tilde Y}_4$ by 
$(A^{I}, \tilde A_{I}, B^{I}, \tilde B_{I})$, where
$I=0,\ldots,h^{(1,1)}(Y_3)$.
The $(A^{I}, \tilde A_{I})$ are those homology cycles that correspond
via mirror symmetry to  the elements  of the  vertical  primary subspace of
$Y_4$ which are obtained by joining the even-dimensional cycles of the
threefold-fibre with the zero-cycle of the base. 
Analogously, $(B^{I}, \tilde B_{I})$
are the cycles which are related to  the elements of the  vertical primary
subspace which are obtained by joining the even-dimensional 
cycles of the threefold with the
two-cycle of the base \cite{M}. 
As noted in \cite{M} 
in the large base limit at leading order
the cycles $(A^{I}, \tilde A_{I})$
 all have vanishing 
intersections with each other and the
same is true for the $(B^{I}, \tilde B_{I})$ cycles. 
The only 
non-vanishing intersections are between $A$-cycles and $B$-cycles and 
the intersection form is given by that of the $Y_3$-fibre.
For a certain choice of cycles and in 
terms of the Poincar\'e dual forms 
$(a^I, \tilde a_I, b^I, \tilde b_I) $
this amounts to
\be\label{isec}
\int_{\tilde Y_4} a^I \wedge\tilde b_J\ =\ \delta^I_J\ =\ 
- \int_{\tilde Y_4} \tilde a_J \wedge b^I\ 
\ee
with all other intersection pairings vanishing.

In order to evaluate the superpotential
$W(\tilde Y_4)$ using eq.\ (\ref{W})
we expand the 4-form flux on $\tilde Y_4$ in this basis
\beqn\label{Fexp}
\frac{\tilde F_4}{2\pi} = \mu_I a^I - \tilde \mu^I \tilde a_I 
+ \nu_I b^I - \tilde \nu^I \tilde b_I, \qquad (\mu_I, \tilde \mu^I,  
\nu_I, \tilde \nu^I)\in \mathbb Z\ .
\eeqn
(We denote it by 
$\tilde F_4$ in order to distinguish it 
from the 4-form flux on $Y_4$.)
The fluxes are not all independent but have to obey
the consistency condition (\ref{ccond}).
This implies
\be\label{fcond}
\frac{1}{24}\,  \chi(\tilde Y_4) = n + 
(\nu_I\tilde\mu^I - \mu_I\tilde\nu^I)\ .
\ee

Inserting (\ref{Fexp}) into (\ref{W}) we arrive at
\be\label{spb}
W= \mu_I \int_{A^{I}}\Omega  
\ -\ \tilde \mu^I \int_{\tilde A_{I}} \Omega
\ +\ \nu_I \int_{B^{I}}\Omega\ -\ \tilde \nu^I \int_{\tilde B_{I}}\Omega \ .
\ee
In order to evaluate the period integrals 
$\int\Omega$  on $\tilde Y_4$ 
one has to note that they are 
mapped via mirror 
symmetry to the periods on $Y_4$, which give to leading order in the 
large volume limit
and in special coordinates the classical volumes of the corresponding 
cycles. These leading terms
in general get quantum corrections. In the large base limit 
the corrections from the base are suppressed and the quantum 
corrections only arise from the threefold fibre. Thus the periods on 
the fourfold are given by those of the threefold, 
multiplied by the classical volume $t^V$ of the base for those cycles which 
contain the base \cite{M}. 
More specifically one has\footnote{We thank P.\ Mayr
for a clarifying discussion concerning the period
integrals.}
\beqn \label{zG}
\int_{A_{I}}\Omega &=&  (1,t^i)  
+\  {\cal O}(e^{i t^V})\ , \qquad 
\int_{B_{I}}\Omega =  t^V (1,t^i)  + \
{\cal O}((t^V)^0) + {\cal O}(e^{i t^V})\ ,
\\
\int_{\tilde A_{I}}\Omega &=&
({\cal{F}}_i, {\cal{F}}_0)
+ {\cal O}(e^{i t^V})\ ,\qquad 
\int_{\tilde B_{I}}\Omega =
t^V ({\cal{F}}_i, {\cal{F}}_0)
+ \ {\cal O}((t^V)^0) + {\cal O}(e^{i t^V})
\ ,\nonumber
\eeqn
where $i=1,\ldots,h^{(1,1)}(Y_3)$.
The vector $\Pi=(1,t^i, {\cal{F}}_i, {\cal{F}}_0)$
corresponds
to the periods of the threefold with
${\cal{F}}_i=\partial_{t^i} {\cal{F}}$
and ${\cal{F}}_0 = 2 {\cal{F}} 
- t^i {\cal{F}}_i$. The $N=2$ prepotential  
is given by \cite{HKT}
\beqn\label{Fdef}
{\cal{F}}&=& {\cal{F}}_{\rm pol} -\frac{1}{(2\pi)^3}\sum_{\{d_i\}}
n_{\{d_i\}} Li_3(e^{2 \pi i \sum t^id_i}) \ ,
\eeqn
where
\beqn\label{Lidef}
{\cal{F}}_{\rm pol}=\frac{1}{6} d_{ijk} t^i t^j t^k  
+ b_i t^i + \frac{1}{2} c \ ,
\qquad
 Li_3(x) \equiv \sum_{j=1}^{\infty} \frac{x^j}{j^3}\ .
\eeqn
The $d_{ijk}$ are the classical 
 intersection numbers of the threefold and the
coefficients 
$b_i,c$ are given in \cite{HKT}. The 
$d_i$ are
 the instanton numbers of the $i$-th $(1,1)$-form,
$d_i=\int_{\cal{C}}e_i$, while 
$n_{\{d_i\}} $ is  the number of
isolated  holomorphic curves $\cal{C}$ of multi-degree  
$(d_1,\ldots
  ,d_{h^{(1,1)}})$ in the threefold fibre and the sum over $j$ takes into
  account multiple coverings.

Inserting (\ref{zG}) into (\ref{spb})
and assuming that the $\mu_I$ and  
$\tilde \mu^I$ are large
so that the ${\cal O}((t^V)^0)$ term
can be neglected we arrive at
\be \label{wint}
 W(\tilde Y_4) 
 =  \mu_0+\mu_i t^i-{\tilde\mu}^i{\cal{F}}_i
-{\tilde \mu}^0{\cal{F}}_0 
+\nu_0 t^V + \nu_i t^it^V-{\tilde\nu}^i{\cal{F}}_it^V 
- {\tilde\nu}^0{\cal{F}}_0t^V\ .
\ee
For future reference let us recall that the 
same period integrals (\ref{zG}) 
have been used in \cite{M}
to derive the K\"ahler potential in the large
base limit to be
\be \label{2dk2a}
K^{(2)}_{IIA} = -\ln [\int_{\tilde Y_4} \Omega \wedge \bar \Omega ] = 
-\ln \big[ (t^V - \bar t^V) 
(2 ({\cal{F}} - \bar {\cal{F}}) - 
(t^i - \bar t^i)({\cal{F}}_i + \bar {\cal{F}}_i))\big]\ .
\ee

$W$ can be written in a more suggestive way
by expressing it not in terms
of special coordinates $t^i$ but rather
in terms of the homogeneous coordinates $X^I$ \cite{HKT}.
These coordinates are commonly used 
in $N=2$ supergravity and are 
holomorphic functions of the special 
coordinates $X^I(t^i)$. Furthermore, one has a
prepotential
$F(X)$ which is a homogeneous function of the $X^I$
of degree two.
The special coordinates are just the particular
coordinate choice 
$t^0 = X^0/X^0 = 1, t^i = X^i/X^0, {\cal F}(t^i) = (X^0)^{-2} F(X)$.
In homogeneous coordinates the periods of the threefold
are
\be \label{generalpi}
\Pi =  (X^I, F_I) 
= X^0 (1,t^i, {\cal{F}}_i, {\cal{F}}_0)\ ,
\ee
while the K\"ahler potential reads 
\be \label{k2aX}
K^{(2)}_{IIA} = -\ln \big[ (t^V - \bar t^V) 
(\bar X^I F_I - X^I \bar F_I)\big]\ .
\ee
This coincides with the K\"ahler potential
given in (\ref{2dk2a}) up to a 
K\"ahler transformation that
amounts to a different 
normalization of the $(4,0)$-form $\Omega$.
Inserting (\ref{generalpi}) into (\ref{wint})
finally yields 
\be \label{wty}
W(\tilde Y_4) = \alpha_I X^I - \beta^I F_I\ ,
\ee
where we abbreviated $\alpha_I = \mu_I + t^V \nu_I$, 
$\beta^I = \tilde \mu^I + t^V \tilde \nu^I$
and discarded an $X^0$-factor by the same K\"ahler
transformation.

Curiously the superpotential (\ref{wty}) coincides
with the superpotential derived in ref.\ \cite{TV}.
It arises in type IIB compactifications on a Calabi-Yau 
threefold with non-vanishing RR- and NS  3-form fluxes 
studied in refs.\ \cite{JM,TV,M2}. 
In this case the type IIB dilaton plays the
role of $t^V$ in (\ref{wty}).
It also is very closely related to the BPS-mass formula
studied in refs.\ \cite{CDFP,CLM} and the 
entropy formula of $N=2$ black holes \cite{FK}. 
This fortunate coincidence saves us from a
detailed analysis of the supersymmetric minima of 
(\ref{wty}) and we can simply refer the reader to
refs.\  \cite{JM,TV,M2,CKLT,CLM}.
One finds that for generic fluxes no supersymmetric
vacuum exists. However, if the 
$\alpha_I,\beta^I$ are appropriately chosen
supersymmetric ground states can exists. 
This can happen if the fluxes are aligned
with cycles of the threefold which can
degenerate at specific points in the moduli space 
\cite{PS,JM,TV,M2,CKLT}.
These points (or subspaces) then correspond to
supersymmetric ground states. They also coincide with
the supersymmetric attractor points studied in refs.\
\cite{FK}. Note that in ref.\ \cite{TV} the
consistency of the compactification
required $\nu_I\tilde\mu^I - \mu_I\tilde\nu^I=0$
while in our case this is replaced with
the generalized condition given in eq.\ (\ref{fcond}).

Before we evaluate $\tilde W (Y_4)$ let us briefly discuss the
symmetry properties of $W(\tilde Y_4)$ as obtained in 
eq.\ (\ref{wty}).
In homogeneous coordinates the period vector $\Pi$
transforms as a symplectic vector according to 
\be\label{symple}
\left( \begin{array}{c}
           F_I \\
           X^I
       \end{array} \right) 
\rightarrow 
\left( \begin{array}{cc}
           A & B \\
           C & D 
       \end{array} \right)
\left( \begin{array}{c}
           F_I \\
           X^I
       \end{array} \right)\ , \qquad 
\left( \begin{array}{cc}
           A & B \\
           C & D 
       \end{array} \right) \in {\rm Sp}(2h^{(1,1)} + 2, \mathbb{Z})\ .
\ee
This transformation leaves the symplectic product
$(\bar X^I F_I - X^I \bar F_I)$ invariant
and via eq.\ (\ref{k2aX}) also the K\"ahler
potential (in homogeneous coordinates). 
The fact that $(F_I, X^I)$ transform as a 
symplectic vector implies that also $(\tilde a_I, a^I)$ and 
$(\tilde b_I, b^I)$ transform according to (\ref{symple}). 
Indeed one verifies that 
the intersection matrix given in (\ref{isec}) is 
left invariant if $(\tilde a_I, a^I)$ as well as 
$(\tilde b_I, b^I)$ transform as symplectic vectors.
Since $\tilde F_4$ has to be symplectically
invariant one infers from (\ref{Fexp}) that 
in turn the fluxes $(\mu_I, \tilde \mu^I)$ 
and $(\nu_I, \tilde \nu^I)$ have to transform as 
symplectic vectors, 
i.e. according to (\ref{symple}).  
Since $t^V$ is invariant we conclude
that $(\alpha_I, \beta^I)$ form a symplectic vector
and $W$ of (\ref{wty}) is invariant.


Having derived the expression (\ref{wint})
for $ W(\tilde Y_4)$ we can use 
eq.\ (\ref{weqw}) to determine the quantum corrections
of $\tilde W(Y_4)$. 
By matching the classical terms of $\tilde W(Y_4)$
as given in eq.\ (\ref{Wtilde}) with the 
classical part of (\ref{wint}) using
(\ref{Fdef}), (\ref{Lidef}) 
one is led to the identification
\beqn 
& &-\tilde\nu^0 {\cal{F}}_0 t^V=\frac{1}{2\pi} \int_{Y_4}t\wedge t \wedge t \wedge
t \ F_0+\mbox{quantum corrections}\ , \\
& &-\tilde\mu^0{\cal{F}}_0-\tilde\nu^i{\cal{F}}_it^V=\frac{1}{2\pi}
\int_{Y_4}t\wedge t\wedge t\wedge F_2+\mbox{q.c.}\ , \\
\label{thrp}& &\nu_it^it^V-\tilde\mu^i{\cal{F}}_i=\frac{1}{2\pi} \int_{Y_4} t\wedge t\wedge
F_4+\mbox{q.c.}\ , \\
\label{twop}& &\nu_0 t^V+\mu_it^i=\frac{1}{2\pi} \int_{Y_4} t\wedge F_6\ ,  \\
\label{onep}& &\mu_0 =\frac{1}{2\pi} \int_{Y_4}F_8=\mbox{const.}\  .
\eeqn 
The right hand sides of  (\ref{twop}), (\ref{onep}) are two- and one-point
functions in the topological A-model and therefore do not
receive instanton corrections \cite{GMP,DVV}. 
{}From eq.\ (\ref{thrp}) using (\ref{Fdef}), (\ref{Lidef}) 
we learn  that the term including ${\cal F}_i$
has a polynomial piece and instanton corrections
while the second term $\nu_i t^i t^V$  is purely
classical. 
 This can also be understood by considering the 
corresponding correlation functions in
the A-model. Let us now go through this computation in
more detail.

\subsubsection{
The superpotential $\tilde W$ generated by 4-form flux} \label{D4}

 We denote the observables of the A-model by 
${\cal O}^{(k)}_M\in H^{(k,k)}(Y_d)$, where  the index
takes the values 
$M=1,\ldots,h^{(k,k)}_V(Y_d)$.
 The two-point functions 
\beqn
\eta^{(k)}_{MN}
=\langle{\cal O}_M^{(k)}{\cal O}_N^{(d-k)}\rangle
=\int_{Y_d} {\cal
  O}_M^{(k)}\wedge {\cal O}_N^{(d-k)}
\eeqn
get no instanton contributions and define a flat metric on the vertical
primary cohomology \cite{DVV}.\footnote{Note that 
this metric 
is not the Zamolodchikov metric 
$G_{A\bar B}= \partial_A\partial_{\bar B} K$
which does receive quantum corrections
as can be seen from eq.\ (\ref{2dk2a}). }
On the other hand the three-point functions 
\beqn\label{3a}
Y_{KLM}^{(k)}=\langle{\cal O}_K^{(1)}{\cal O}_L^{(k)}{\cal O}_M^{(d-k-1)}\rangle, 
\eeqn
do
receive instanton corrections. Because of their factorization
properties all other amplitudes can be
expressed in terms of the two-  and three-point functions $\eta^{(k)}_{MN},
Y_{KLM}^{(k)}$ \cite{GMP}. 

Choosing the 4-form flux as
\be
\frac{F_4}{2\pi}=\sum_{N=1}^{h^{2,2}_V}\lambda^N{\cal O}_N^{(2)}\ ,
\ee 
 ref.\ \cite{L}  proposed
the following formula
\be\label{wolli}
\partial_{t^A}\partial_{t^B}\tilde W(Y_4)=\sum_N
\lambda^{N}\langle{\cal O}^{(1)}_A{\cal
  O}^{(1)}_B{\cal O}^{(2)}_N\rangle \ .
\ee
In the following we evaluate  this
three-point function for a threefold-fibred fourfold in the large base limit and show that
(\ref{wolli}) is consistent with (\ref{thrp}). 

Let us first consider the part of $F_4$
which has one component
 in the base 
\beqn \label{G}
 \frac{F_4}{2\pi}&=& \sum_{N=1}^{h^{1,1}(Y_3)}\lambda^N{\cal O}_N^{(2)}=\sum_{i=1}^{h^{(1,1)}(Y_3)}\lambda^{Vi}e_V\wedge e_i, 
\eeqn
where $\lambda^{Vi}$ is large.
 In the large base
 limit, there are no instanton corrections from the base so that according to
 the classical intersection numbers $e_V$ occurs
 at most once in any correlation function. The divisor which is dual to $e_V$
 is the threefold-fibre and projects the amplitude of the fourfold to the
 threefold. In particular, for the three-point function with the 4-form flux
 as in (\ref{G}), the three-point function on the fourfold is projected to the
 three-point function on the threefold \cite{M}
\be \label{yijk}
\partial_{t^i}\partial_{t^j}\tilde W(Y_4)=
\sum_N \lambda^N\langle{\cal
  O}^{(1)}_i{\cal O}^{(1)}_j{\cal O}^{(2)}_N\rangle_{Y_4}
=\sum_k  \lambda^{Vk}\langle{\cal
  O}^{(1)}_i{\cal O}^{(1)}_j{\cal O}^{(1)}_k\rangle_{Y_3}
=\sum_k \lambda^{Vk}Y_{ijk}, 
\ee
where 
\beqn
  {\cal O}^{(2)}_N= e_V\wedge e_k\ , \qquad
{\cal O}^{(1)}_k=e_k\ . 
\eeqn 
In special coordinates the three-point
 function $Y_{ijk}$ on $Y_3$  is the third derivative of a holomorphic
 prepotential \cite{HKT}
\beqn \label{fijk}
 Y_{ijk}={\cal F}_{ijk}, \qquad {\cal F}_{ijk}
=\partial_{t^i}\partial_{t^j}\partial_{t^k}{\cal F}\ .
\eeqn
Inserting (\ref{fijk}) in eq.\ (\ref{yijk})
we learn that 
$\partial_{t^i}\partial_{t^j}\tilde W(Y_4)$
can be integrated and indeed coincides with
the instanton corrected part of the expression
(\ref{thrp}).

In order to get the full superpotential, 
we  still have to consider the part with
the 4-form flux restricted  to the
threefold fibre, i.e.
\beqn 
 \frac{F_4}{2\pi}&=&\sum_{i=1}^{h^{1,1}(Y_3)}\lambda^i{\cal O}_i^{(2)}.
\eeqn
We have to distinguish two cases. First we consider the three-point function
which contains one observable 
${\cal  O}^{(1)}$ corresponding to the base.
In this case the three-point function of the fourfold is projected to the 
two-point function of the threefold fibre:
\beqn
\partial_{t^V}\partial_{t^j}\tilde W(Y_4)=\lambda^{i}\langle{\cal O}^{(1)}_V{\cal O}^{(1)}_j{\cal O}^{(2)}_i\rangle_{Y_4}=\lambda^{i} \langle{\cal O}^{(1)}_j{\cal O}^{(2)}_i\rangle_{Y_3}=\lambda^{i}\eta^{(1)}_{ji}.
\eeqn 
As already mentioned above, two-point functions receive no worldsheet instanton
corrections and 
the classical part of the amplitude is already the exact expression. 
Integrating twice we obtain the term in (\ref{thrp}) which does not contain
instanton corrections. 

Finally, the contribution to the superpotential which has no component in the base is
subleading in the limit where $\lambda^{Vi}$
and $t^V$ are large.
To summarize, we confirmed the expression
given in eq.\ (\ref{thrp}) by considering 
correlation functions in the topological A-model,
that is without using mirror symmetry.
Altogether we thus have
\beqn
\tilde W(Y_4)= \lambda^{k}\eta^{(1)}_{jk}t^jt^V+\lambda^{Vk}{\cal F}_k,
\eeqn
with the relations $\lambda^{k}\eta^{(1)}_{jk}=\nu_j,
\lambda^{Vk}=-\tilde\mu^k$.

\subsection{The superpotential for $K3$-fibred fourfolds}

In this section 
we further specify to $K3$-fibred fourfolds
so that we are able to use
the type IIA -- heterotic duality and derive
the heterotic superpotential including
the type IIA quantum corrections.
Specifically we choose the fourfold to be
$K3$-fibred over a large base $\mathbb{F}_n$
as we did in section \ref{hetiia}. 
Such fourfolds can also be viewed
as $Y_3$ fibred fourfolds over  a base $\mathbb{P}_1$
where the $Y_3$-fibre is itself $K3$-fibred
over a second $\mathbb{P}_1$.
This allows us to make contact with the results
of the last section and merely specify them
for $K3$-fibred threefolds.

The case of $K3$-fibred threefolds has been 
studied in detail in connection with
the heterotic -- type IIA duality for
$N=2, D=4$ string vacua \cite{D4rev}.
For such threefolds the prepotential ${\cal F}$
obeys in the large $\mathbb{P}_1$ limit 
\be\label{FK3}
{\cal F} = t^U \eta_{{\hat \imath}{\hat \jmath}}
t^{\hat \imath} t^{\hat \jmath} 
+ {\cal F}^{(1)}(t^{\hat \imath}, t^{\hat \jmath} )
+ {\cal O}(e^{it^U})
\ ,
\ee
where $t^U$ is the modulus parameterizing 
the $\mathbb{P}_1$ base while the $t^{\hat \imath}$
denote the moduli of the $K3$-fibre
except those from reducible bad fibres.
($\eta_{{\hat \imath}{\hat \jmath}}$ has been defined
in (\ref{etadef}).)

Inserting (\ref{FK3}) into  (\ref{wint})
one obtains the superpotential for 
$K3$-fibred fourfolds in the large $\mathbb{F}_n$ limit.
If one assumes that the duality map
given in (\ref{abb}) continues to hold
in the presence of background fluxes
one can use it to derive a heterotic superpotential
$W_{\rm het}(\tau,\rho,n^a,u,v)$.
However, this might be possible only for
a subset of the fluxes considered so far \cite{LLP}.
 
As we discussed in section 2.3 both
the heterotic vacua and the dual type IIA vacua
enjoy a natural action
of the symmetry group
$SL(2,\mathbb Z)\times SL(2,\mathbb Z)$.
This is a subgroup of the 
symplectic $Sp(2h^{(1,1)}+2)$ symmetry group 
which we discussed in section \ref{Wsec}.
More concretely the transformation (\ref{sl2z}) acts as a 
specific symplectic transformation (\ref{symple}) 
on the homogeneous coordinates corresponding 
to $(\rho, \tau, n^a, u)$ respectively
$t^i$, which in the large base limit 
leaves $u$ and $t^U$ invariant \cite{DKLL}.
{}From the discussion of section \ref{Wsec} it is 
clear that the 
$SL(2,\mathbb Z)\times SL(2,\mathbb Z)$ transformations
leave both the K\"ahler potential and 
the superpotential invariant 
if they are expressed in terms of homogeneous coordinates. 
In special coordinates the $SL(2,\mathbb Z)\times SL(2,\mathbb Z)$ 
transformations 
instead lead to a K\"ahler transformation of the 
K\"ahler- and the superpotential which also leaves the 
action invariant. 


Finally we remark that the heterotic origin of some 
of the background fluxes $F_2$ and $F_4$ might be 
understood from the 
duality between the heterotic string on $T^4$ and the 
type IIA theory on $K3$, as was first noted in \cite{PS}. 
For example, denoting by $e_{U/V}$ the two $(1,1)$-forms 
stemming from the base $\mathbb{F}_n$ and by $e_{\hat \imath}$ those 
of the fibre $K3$, a background $F_4 \sim \lambda^{U/V \hat \imath}
e_{U/V} \wedge e_{\hat \imath}$ 
can be interpreted as giving a background flux along $e_{U/V}$ to the 
field strength of the $D=6$ vectors coming from expanding $A_3$ 
in terms of the $e_{\hat \imath}$. These vectors have heterotic
counterparts $A^{\hat \imath}$ according to the $D=6$ duality. 
Expanding their field strength as $F^{\hat \imath}_2
\sim \lambda^{U/V \hat \imath} e_{U/V}$ should correspond 
to the above background $F_4$. 
A similar situation holds for 
$F_4 \sim \lambda^{UV} e_{U} \wedge e_{V}$ and 
$F_2 \sim \lambda^{U/V}e_{U/V}$, 
but there seems to be no such simple heterotic correspondence 
for all other background fluxes. 
Some of them might be related to non-vanishing torsion
on the heterotic side \cite{DRS}.

\vskip 2cm

\noindent
{\Large {\bf Acknowledgements}}

We thank A.\ Klemm, P.\ Mayr, H.\ Nicolai and H.\ Singh 
for usefull conversations and correspondence.
We also thank the authors of ref.\ \cite{CKLT}
for communicating their results prior to
publication.

This work is supported by
DFG (the German Science Foundation), 
GIF (the German--Israeli 
Foundation for Scientific Research),
DAAD (the German Academic Exchange Service)
and the Landesgraduierten\-f\"orderung
Sachsen-Anhalt.


\vspace{1cm}
\appendix
\noindent
{\Large {\bf Appendix}}

\setcounter{equation}{0}
\setcounter{section}{0}

\section{Notation}
\label{notation}

The signature of the space-time metric is $(-+ \ldots +)$.
The Levi-Civita symbol is defined to transform as a 
tensor, i.e. we have 
\be
\epsilon^{1 \ldots D} =  (\pm g^{(D)})^{-1/2} \qquad 
\mbox{and} \qquad 
\epsilon_{1 \ldots D} = \pm (\pm g^{(D)})^{1/2}\ ,
\label{eqepsilon}
\ee
where the $+$ sign corresponds to Euclidean and the $-$ sign  to Minkowskian signature.
Our conventions for the Riemann curvature tensor 
are
\be
R^{\mu}_{\nu \rho \sigma} = \partial_{\rho} 
\Gamma^{\mu}_{\nu \sigma} - \partial_{\sigma} 
\Gamma^{\mu}_{\nu \rho} + \Gamma^{\omega}_{\nu 
\sigma} \Gamma^{\mu}_{\omega \rho} - 
\Gamma^{\omega}_{\nu \rho} 
\Gamma^{\mu}_{\omega \sigma}\ ,
\label{eqcurvature}
\ee
where we use the following definition of the 
Christoffel symbols:
\be
\Gamma^{\mu}_{\nu \rho} = \frac{1}{2} g^{\mu \sigma} 
(\partial_\nu g_{\sigma \rho} + \partial_\rho 
g_{\sigma \nu} - \partial_\sigma g_{\nu \rho})\ .
\label{eqchrisis}
\ee
The Ricci tensor is defined as
\be
R_{\mu \nu} = R^{\rho}_{\mu \rho \nu}\ .
\label{eqricci}
\ee
We are thus using the (+++) conventions of \cite{MTW}.

Furthermore a $p$-form $A_p$ can be expanded as
\be
A_p = \frac{1}{p!} A_{\mu_1 \ldots \mu_p} 
d x^{\mu_1} \wedge \ldots \wedge d x^{\mu_p}\ ,
\label{eqpform}
\ee
which has an obvious generalisation to $(p,q)$-forms:
\be
A_{p,q} = \frac{1}{p! q!} A_{i_1 \ldots i_p 
\bi_1 \ldots \bi_q} d \xi^{i_1} \wedge \ldots \wedge 
d \xi^{i_p} \wedge d \bar{\xi}^{\bi_1} \wedge \ldots 
\wedge d \bar{\xi}^{\bi_q}\ . 
\label{eqpqform}
\ee
The exterior derivative is defined as
\be
dA_p = \frac{1}{p!} \pu A_{\mu_1 \ldots \mu_p} d x^\mu \wedge
 d x^{\mu_1} \wedge \ldots \wedge d x^{\mu_p}\ , 
\label{eqextderiv}
\ee
which entails because of 
\be
F_{p+1} = dA_p = \frac{1}{(p+1)!} F_{\mu_1 \ldots \mu_{p+1}} 
d x^{\mu_1} \wedge \ldots \wedge d x^{\mu_{p+1}}
\label{eqfstr}
\ee   
the relation
\be
F_{\mu_1 \ldots \mu_{p+1}} = (p+1) \partial_{[\mu_1} 
A_{\mu_2 \ldots \mu_{p+1}]}\ .
\label{eqfieldstr}
\ee
With this definition the action for a $p$-form 
potential $A_p$ is given by
\be
-\frac{1}{4} \int d^D x \sqrt{-g^{(D)}} |F_{p+1}|^{2} = 
-\frac{1}{4} \int d^D x \frac{\sqrt{-g^{(D)}}}{(p+1)!} 
F_{\mu_1 \ldots \mu_{p+1}} F^{\mu_1  \ldots \mu_{p+1}}\ . 
\label{eqpformaction}
\ee 
The Hodge star operator for a $p$-form is defined as
\be
\star A_p = \frac{1}{p!(D-p)!} A_{\mu_1 \ldots \mu_p} 
\mbox{$\epsilon^{\mu_1 \ldots \mu_p}$}_{\nu_{p+1} \ldots 
\nu_D} dx^{\nu_{p+1}} \wedge \ldots \wedge dx^{\nu_D}\ .
\label{eqstar}
\ee
It is generalised to $(p,q)$-forms on a Hermitian 
manifold (where $D$ now denotes its complex dimension) 
by
\beqn
\star A_{p,q} = \frac{(-1)^{(D-p) q +1/2 D(D-1)} i^D}
{p! q! (D-p)! (D-q)!} 
\bar{A}_{i_1 \ldots i_q \bi_1 \ldots \bi_p} 
\mbox{$\epsilon^{i_1 \ldots i_q}$}_{\bj_{q+1} \ldots 
\bj_D} \mbox{$\epsilon^{\bi_1 \ldots \bi_p}$}_
{j_{p+1} \ldots j_D} \non
\times 
d \xi^{j_{p+1}} \wedge \ldots \wedge d \xi^{j_D} \wedge 
d \bar{\xi}^{\bj_{q+1}} \wedge \ldots \wedge 
d \bar{\xi}^{\bj_{D}}\ ,
\label{eqstarcomplex}
\eeqn
where $\bar{A}_{i_1 \ldots i_q \bi_1 \ldots \bi_p} = 
\overline{A_{i_1 \ldots i_p \bi_1 \ldots \bi_q}}$. 
This definition ensures that we have the following 
expression for the scalar product of two $(p,q)$-forms 
on a Hermitian manifold:
\beqn
(A_{p,q},B_{p,q}) &  \equiv & \int A_{p,q} \wedge \star B_{p,q} \non
& = &\frac{(-1)^{1/2 D(D-1)} i^D}{p! q!} \int 
\sqrt{g^{(D)}} A_{i_1 \ldots i_p \bi_1 
\ldots \bi_q} \bar{B}^{\bi_1 
\ldots \bi_q i_1 \ldots i_p} d^D \xi d^D 
\bar{\xi}\ .  
\label{scalarprod}
\eeqn


\section{$T^2$ compactification of $D=4$ 
super\-gravity}
\label{hetred}
\setcounter{equation}{0}

In this appendix we give some details of the 
$T^2$ reduction of the $D=4,N=1$ supergravity
given in equation (\ref{het4d}). As already mentioned 
in the main text there are two equivalent ways to 
derive the effective action in $D=2$. One can either
directly compactify the $D=4$ action on a torus. This leads naturally to the
variables relevant for the purpose of this paper. 
One can also perform the reduction in two steps with the 
detour over $D=3$. For completeness we give this derivation too. In this case one can make use of the 
results given in \cite{HL}.  

Let us first perform the direct torus-reduction. In contrast 
to the main text we here work in the Einstein-frame because
this is easier to handle in the general case where one does not 
specify the K\"ahler potential and gauge kinetic function 
to their tree level form. Inserting 
(\ref{ahat}) and (\ref{gmn}) (but now in the Einstein-frame) 
into (\ref{het4d}), using the 
definitions (\ref{tau}), (\ref{dhz}) and (\ref{dmuna}) 
one derives the following $D=2$ effective Lagrangian
\beqn \label{l2het}
{\cal L}_{\rm het}^{(2)} & = & \sqrt{h_{\se}} \left(- \frac{1}{2} \pu \po \sigma
+ \frac{\pu \tau \po \bar \tau}{(\tau - \bar \tau)^2}
- G_{I \bar J}^{(4)} \pu \Phi^I \po \bar \Phi^{\bar J} 
\right) \\
& & -  (\tau - \bar \tau)^{-1}\left(i ({\rm Re} f_{ab}) 
D_\mu n^a D^\mu\bar n^b 
- \frac{1}{2} \epsilon^{\mu \nu} 
(\pu {\rm Im} f_{ab}) \left( \bar n^a D_\nu n^b - n^a D_\nu \bar n^b \right)\right),
\nonumber
\eeqn
where $h_{\se}$ denotes the determinant of the 
torus metric in the Einstein-frame and we have chosen the 
conformal gauge for the two-dimensional space-time metric 
\be
g^{(2)}_{\mu \nu} = e^\sigma h^{1/4}_{\se} \eta_{\mu \nu}.
\ee 
As in equation (\ref{het4d}) the $\Phi^I$ are all moduli 
of the $D=4$ Lagrangian including the dilaton $S$. 
Furthermore we have omitted all vectors because they do not 
have dynamical degrees of freedom in $D=2$.

As has been shown in \cite{GHR} the moduli space 
of non-linear $(2,2)$ sigma-models in $D=2$ 
is in general not K\"ahler, when chiral 
and twisted chiral multiplets are present. 
Nevertheless the Lagrangian can be expressed 
by second-order partial derivatives of a real function 
of the moduli, analogous to the K\"ahler 
potential of a K\"ahler manifold.   
In case of dilaton-supergravity the general form 
of the Lagrangian has been given in \cite{GGW}.
In our case  $\sqrt{h_{\se}}$ takes
the role of the $D=2$ dilaton.\footnote{In fact $\sqrt{h_{\se}}$ is the 
heterotic $D=2$ dilaton because $e^{-2 \dhz} = e^{-2 \dhv} \sqrt{h} = 
\sqrt{h_{\se}}$. The last equality is due to the Weyl
rescaling relating 
the $D=4$ Einstein- and string-frame metrics.} 
The Lagrangian (\ref{l2het}) is the sum of 
two terms, one multiplied by $\sqrt{h_{\se}}$, the other one not. 
Both terms can seperately be expressed via a potential.

Defining the two real functions 
\beqn
K_1^{\rm het} & = & K^{(4)}(\Phi, \bar \Phi) + \ln(-i (\tau - \bar \tau)), \\
K_2^{\rm het} & = & \frac{i}{2} ({\rm Re}f_{ab}(\Phi)) 
\left[ (\tau - \bar \tau)^{-1} (n^a - \bar n^a) (n^b - \bar n^b) \right]
\eeqn
and denoting collectively $(\tau, n^a)$ as $\chi^\Sigma$ 
we can express (\ref{l2het}) in the form 
\beqn\label{2dhet}
{\cal L}_{\rm het}^{(2)} & = & \sqrt{h_{\se}} \left(- \frac{1}{2} \pu \po \sigma
- \big(\partial_{\Phi^I} \bar{\partial}_{\bar \Phi^{\bar J}} K_1^{\rm het}\big) \, \pu \Phi^I 
\po \bar \Phi^{\bar J} + \big(\partial_{\chi^\Sigma} 
\bar{\partial}_{\bar \chi^{\bar \Lambda}} K_1^{\rm het}\big) \, \pu \chi^\Sigma 
\po \bar \chi^{\bar \Lambda} \right) \non
& & \mbox{} + \big(\partial_{\chi^\Sigma} \bar{\partial}_{\bar \chi^{\bar \Lambda}}
K_2^{\rm het}\big) \,
 \pu \chi^\Sigma \po \bar \chi^{\bar \Lambda} \\
& & \mbox{} - \epsilon^{\mu \nu} \big(\bar{\partial}_{\bar \Phi^{\bar J}}
\partial_{\chi^\Sigma} K_2^{\rm het}\big) \, \pu \bar \Phi^{\bar J} 
\partial_\nu \chi^\Sigma - \epsilon^{\mu \nu}
\big(\partial_{\Phi^I} \bar{\partial}_{\bar \chi^{\bar \Lambda}} K_2^{\rm het}\big) \, 
\pu \Phi^I \partial_\nu \bar \chi^{\bar \Lambda}. \nonumber
\eeqn
This is of the form given in \cite{GHR,GGW}.
Obviously the moduli space is not K\"ahler. As explained in \cite{GHR} 
the terms proportional to $\epsilon^{\mu \nu}$ always combine a derivative 
with respect to a chiral field with a derivative with respect to a 
twisted chiral field. The $\Phi^I$ stem from the chiral $D=4$ moduli 
and continue to be chiral in $D=2$. In view of the last line in  
(\ref{2dhet}) this means that $(\tau, n^a)$ reside in twisted 
chiral multiplets. 
  
In the alternative derivation of the $D=2$ effective action one 
reduces the $D=3$ effective action obtained in \cite{HL} on 
a further circle. Recall that in $D=3$ an Abelian vector is dual to
a scalar and thus a vector multiplet can
be dualized to a scalar multiplet.
Hence the three-dimensional Lagrangian on the 
Coulomb branch can be entirely expressed
in terms of chiral multiplets.
The bosonic part of this Lagrangian
is given in the Einstein-frame by 
\be 
{\cal L}_{\rm het}^{(3)} = \sqrt{g^{(3)}}
\left(\frac{1}{2} R^{(3)} 
- G_{\bar\Lambda\Sigma} \partial_m
\bar{Z}^{\bar\Lambda} 
\partial^m Z^{\Sigma} 
\right)\ ,
\qquad m=0,1,2\ ,
\label{l3}
\ee
where $Z^{\Sigma}$ are the complex
scalar fields in the three-dimensional
chiral multiplets. They
contain the moduli of the four-dimensional heterotic theory
$\Phi^I$, the radius and the Kaluza-Klein gauge boson
of $S^1$ combined into a complex scalar $T$
and finally the scalars $D^a$ which reside in
the chiral multiplets dual to the vector multiplets.\footnote{For the 
exact definition of the fields $T$ and $D^a$ see \cite{HL}.
Note however, that we performed a redefinition 
$T, D^a \rightarrow i T, i D^a$ here.}
Hence we have a decomposition 
$Z^\Sigma = (\Phi^I, T, D^a)$.
Supersymmetry constrains the metric
$G_{\bar\Lambda\Sigma}$ to be K\"ahler, 
$G_{\bar\Lambda\Sigma}
=\bar\partial_{\bar\Lambda}\partial_{\Sigma}K^{(3)}_{\rm
het}$,
and in terms of the
scalars $Z^\Sigma$ the K\"ahler potential reads
\be
K^{(3)}_{\rm het} = 
K^{(4)}_{\rm het}(\Phi,\bar{\Phi}) 
- \ln[-i( T - \bar{T}) + \frac{1}{2} (D - \bar{D})^a 
\R^{-1}_{ab}(\Phi) (D - \bar{D})^b ]\ .  
\label{khet}
\ee

${\cal L}_{\rm het}^{(3)}$ can be further reduced on
a second $S^1$ using the Ansatz
\be 
g^{(3)}_{mn} = \left( \begin{array}{cc}
                 g^{(2)}_{\mu \nu}  & 0\\
                 0 & \rr^2
                 \end{array}
          \right)\ ,
\label{s1red}
\ee
where $\mu,\nu = 0,1$ and $\rr$
is the radius of the $S^1$ measured in the $D=3$ Einstein-frame 
metric. There are no new
Kaluza-Klein gauge bosons in this reduction
since they contain no physical degree
of freedom.
Inserting (\ref{s1red}) into (\ref{l3}) results in
\be 
{\cal L}_{\rm het}^{(2)} = \sqrt{g^{(2)}}\rr[\frac{1}{2} R^{(2)} 
  - G_{\bar \Lambda 
\Sigma} \pu \bar Z^{\bar \Lambda} \po Z^\Sigma]\ .
\label{l2}
\ee
Choosing the conformal gauge
\be
g^{(2)}_{\mu \nu} = e^\sigma \eta_{\mu \nu}\ ,
\ee
and using the relation $r = e^{-2 \dhz}$ between the 
radius of the circle measured in the $D=3$ 
Einstein-frame metric and the 
$D=2$ heterotic dilaton defined in equation (\ref{dhz})
one derives
\be 
{\cal L}_{\rm het}^{(2)} =  e^{-2 \dhz} 
\left[ -\frac{1}{2} \pu \po \sigma - G_{\bar \Lambda 
\Sigma} \pu \bar Z^{\bar \Lambda} \po Z^\Sigma 
\right]\ .
\label{l23}
\ee 
The physical degrees of freedom are exactly
the same as in $D=3$. Also the fact that all scalars $Z^\Sigma$ are 
members of chiral multiplets is inherited from $D=3$. 
Thus in contrast to  equation (\ref{2dhet}) no twisted chiral multiplets occur and the moduli 
space is therefore a K\"ahler manifold.
In fact the sigma-model
geometry is unchanged in the reduction from
$D=3$ to $D=2$, that is 
$ G_{\bar \Lambda \Sigma}$ is the same 
K\"ahler metric with the same K\"ahler 
potential as in $D=3$.


\section{Dimensional reduction of type IIA 
supergravity on Calabi-Yau fourfolds}
\label{IIAred}
\setcounter{equation}{0}

In this appendix we present some of the details 
of the dimensional reduction of the type IIA 
supergravity Lagrangian (\ref{IIA10d}) on Calabi-Yau 
fourfolds including the $(2,1)$-moduli which have been 
neglected in the main text. We follow closely the
procedure applied in ref.\ \cite{BCF,HL} for
compactification of type IIA on Calabi-Yau threefolds.
Starting from the Ansatz
\be
{g}^{(10)}_{MN}(x,y) = \left( \begin{array}{cc}
                     {g}^{(2)}_{\mu \nu}(x) & 0 \\
                     0 & {g}^{(8)}_{ab}(x,y)
                     \end{array} \right),
\label{metricansatz}
\ee
for the D=10 space-time metric one considers infinitesimally
small deformations of the metric 
\be
{g}^{(8)}_{ab} = \hat{g}^{(8)}_{ab}(\langle M\rangle)
+\delta {g}^{(8)}_{ab}(\langle M\rangle, \delta M(x))\ ,
\ee
where $\hat{g}^{(8)}_{ab}$ is a  background 
metric and 
$\delta {g}^{(8)}_{ab}$ its deformation.
Demanding that $\delta {g}^{(8)}_{ab}$
preserves the Calabi-Yau condition
one can expand it  in terms
of non-trivial harmonic forms on $Y_4$.
It is convenient to introduce complex coordinates 
$\xi_j \, (j=1,\ldots, 4)$ for $Y_4$
defining
$\xi_j = \frac{1}{\sqrt{2}} (y_{2j-1} + i y_{2j})$.
For the deformation of the K\"ahler form  
one has\footnote{In the following we 
omit the superscript (8) at the internal metric.}
\be
i \delta g_{i \bj} 
= \sum_{A=1}^{h^{1,1}} \delta M^{A}(x)\, 
e^{A}_{i \bj}\ ,
\label{kaehlerdef}
\ee 
where
$e^{A}$ is an appropriate basis of $H^{1,1}(Y_4)$
and $M^{A}(x)$ are the corresponding real moduli.
For the deformations of the complex structure one has
\be
\delta g_{\bi \bj} = 
\sum_{\alpha=1}^{h^{3,1}} 
\delta {Z}^{ \alpha}(x)\,
{b}^{ \alpha}_{\bi \bj}\ , 
\label{csdef}
\ee 
where ${Z}^{ \alpha}(x)$ are
complex moduli and $b^{\alpha}_{\bi \bj}$ is related to the basis 
$\Phi^{\alpha}$ of $H^{3,1}(Y_4)$
by an appropriate contraction with
the anti-ho\-lo\-mor\-phic 4-form $\bar\Omega$ 
on $Y_4$  \cite{CO}:
\be
b^{\alpha}_{\bi \bj} = - \frac{1}{3 | \Omega 
|^2} \bar{\Omega}_{\bi}^{klm} 
\Phi^\alpha_{klm \bj}\ ,\qquad
| \Omega |^2 \equiv 
\frac{1}{4!} \Omega_{ijkl} \bar{\Omega}^{ijkl}\ .
\label{bphi}
\ee 
The 3-form $C_3$ is expanded in terms of 
the  (2,1)-forms $\Psi^{I}_{ij\bar{k}}$. 
More precisely 
\be
C_{ij\bar{k}} = \sum_{I=1}^{h^{2,1}}
N^{I}(x)\, \Psi^{I}_{ij\bar{k}} \ ,
\qquad
C_{\bi\bj k} = \sum_{I=1}^{h^{2,1}}
 \bar{N}^{\bar{J}}(x)\, 
\bar{\Psi}^{\bar{J}}_{\bi\bj k}\ .
\label{3form}
\ee
Finally the antisymmetric tensor is expanded 
in terms of the (1,1)-forms $e^A$ according to 
\be \label{bij}
B_{i \bj} = \sum_{A=1}^{h^{1,1}}
a^{A}(x)\, e^{A}_{i \bj}\ .
\ee
Again the vectors are neglected because they do
not propagate in $D=2$. 

The dimensional reduction is performed by 
inserting (\ref{metricansatz}), 
(\ref{3form}) and (\ref{bij}) into (\ref{IIA10d}). 
One has to take into account, that the (2,1)-forms 
depend on the complex structure of the 
Calabi-Yau. 
The  basis $\Psi^I$ of $(2,1)$-forms can locally 
be chosen to 
depend holomorphically on the complex structure
or in other words 
\be
\partial_{\bar Z^{\bar \alpha}}\Psi^I = 0\ ,\qquad
\partial_{Z^\alpha}\Psi^I \neq 0 \ .
\ee
The derivative 
$\partial_{Z^\alpha}\Psi^I$ can be expanded into
$(1,2)$- and $(2,1)$-forms with
complex-structure dependent
coefficient functions $\sigma$ and $\tau$
\be \label{dzpsi}
 \partial_{Z^\alpha}  \Psi^I = 
\sigma_{\alpha I K}(Z, \bar{Z}) \,
\Psi^K + 
\tau_{\alpha I \bar{L}}(Z, \bar{Z}) \,
\bar{\Psi}^{\bar{L}}\ .
\ee 

We define a metric 
$G_{I \bar{J}}$ and
intersection numbers $d_{A I \bar{J}}$ 
on the space of $(2,1)$-forms
\beqn\label{G21}
G_{I \bar{J}} &\equiv& \frac{1}{4}
\int_{\footnotesize Y_4} \Psi_I 
\wedge \star {\Psi}_J, \\
d_{A I \bar{J}} & \equiv & \int_{\footnotesize Y_4} e_{A} \wedge
\Psi_{I} \wedge \bar{\Psi}_{\bar{J}}\ .  \nonumber 
\eeqn 
They are related via \cite{HL}
\be
G_{I \bar{J}} = - \frac{i}{2} d_{A I \bar{J}} 
M^A \ . 
\label{g21d}
\ee
Because of (\ref{dzpsi}) they depend on the 
complex structure moduli $Z^\alpha$.
Using (\ref{G21}) and (\ref{gab}) the 
dimensional reduction of (\ref{IIA10d}) yields
\beqn
\label{2a2p}
{\cal L}_{\rm IIA}^{(2)} & = & \sqrt{-g^{(2)}}\,
e^{-2 \dzzw} \left(\frac{1}{2} R^{(2)} 
+ 2 \partial_{\mu} \dzzw \partial^{\mu} \dzzw  - G_{A \bar B} 
\partial_{\mu} t^{A} 
\partial^{\mu} \bar{t}^{\bar B} - G_{\alpha \bar \beta} \partial_{\mu} 
Z^{\alpha} \partial^{\mu} \bar Z^{\bar \beta} \right) \non
& - &  \sqrt{-g^{(2)}}\, \left( 
G_{I \bar{J}} D_{\mu} N^{I} 
D^{\mu} \bar{N}^{\bar{J}} + \frac{1}{4}  d_{A I \bar{J}} 
\epsilon^{\mu \nu} \pu a^{A} \left[ N^{I} D_{\nu} 
\bar{N}^{\bar{J}} - \bar{N}^{\bar{J}}D_\nu N^I \right]\right), 
\eeqn  
where
\be
D_\mu N^I = \pu N^I + N^K \sigma_{\alpha K I} (Z, \bar{Z})
\pu Z^\alpha + \bar{N}^{\bar{L}} 
\bar{\tau}_{\bar{\beta} \bar{L} I} (Z, \bar{Z})
\pu \bar{Z}^{\bar{\beta}}\ .  
\label{Dmu}
\ee
As in the heterotic case there are two terms, one has a factor 
$e^{-2 \dzzw}$ and the other one not. Again both can seperately 
be expressed via a potential. We define 
\beqn
K_1^{\rm IIA} & = & -\ln[\int_{Y_4}\Omega\wedge\bar \Omega] 
+ \ln \vol , \\
K_2^{\rm IIA} & = &  \frac{i}{\sqrt{2}} (t^A - \bar{t}^A) \left( 
\frac{i}{2} d_{AM\bar L} \Gi_{\bar{J} M}^{-1} 
\Gi_{\bar{L} I}^{-1} \nN^I \bar{\nN} \, ^{\bar J}
-\omega_{A I K} \nN^I \nN^K - \bar{\omega}_{A \bar J \bar L}
\bar{\nN} \, ^{\bar J} \bar{\nN} \, ^{\bar L} \right) , \nonumber
\eeqn
where we have introduced new coordinates
\be \label{nhat}
\nN^I = \Gi_{I \bar{J}}
(Z, \bar{Z})\, \bar{N}^{\bar{J}}\ , 
\ee
with
\be \label{ghat}
\Gi_{I \bar{J}} = - \frac{i}{2} d_{A I \bar{J}} c^A  
\ee
and $c^A$ is a constant nonzero vector. 
The $\omega_{A I K}$ are functions of $Z^\alpha$ and 
$\bar{Z}^{\bar{\alpha}}$ which have to obey
\be
\partial_{\bar{Z}^{\bar{\alpha}}} \omega_{A I K} = 
-\frac{i}{4} \Gi^{-1}_{\bar{L} I} \Gi^{-1}_{\bar{J} K} 
d_{A M \bar{L}} 
\bar{{\tau}}_{\bar{\alpha} \bar{J} M}\ ,  
\label{omega}
\ee
but are otherwise unconstrained.

With the help of $K_1^{\rm IIA}$ and $K_2^{\rm IIA}$ 
and denoting collectively 
the fields $(Z^\alpha,\nN^I)$ as $\Phi^\Sigma$ 
the Lagrangian (\ref{2a2p}) can be expressed as
\beqn
\label{2a2b}
{\cal L}_{\rm IIA}^{(2)}& =& e^{-2 \dzzw} \left(-\frac{1}{2} \pu \po \sigma 
- (\partial_{\Phi^\Sigma} \bar{\partial}_{\bar{\Phi}^{\bar{\Lambda}}} 
K_1^{\rm IIA})
\pu \Phi^\Sigma \po \bar{\Phi}^{\bar{\Lambda}} + 
(\partial_{t^A} \bar{\partial}_{\bar{t}^{\bar{B}}} K_1^{\rm IIA}) \pu t^A \po 
\bar{t}^{\bar{B}} \right) \non
&&\mbox{} -(\partial_{\Phi^\Sigma} \bar{\partial}_{\bar{\Phi}^{\bar{\Lambda}}} K_2^{\rm IIA})
\pu \Phi^\Sigma \po \bar{\Phi}^{\bar{\Lambda}} +\epsilon^{\mu \nu} 
(\bar{\partial}_{\bar{\Phi}^{\bar{\Lambda}}} \partial_{t^A} K_2^{\rm IIA}) 
\pu \bar{\Phi}^{\bar{\Lambda}} \partial_\nu t^A\\
&&\mbox{}
+\epsilon^{\mu \nu} (\partial_{\Phi^\Lambda} \bar{\partial}_{\bar{t}^{\bar{A}}} K_2^{\rm IIA})
\pu \Phi^\Lambda \partial_\nu \bar{t}^{\bar{A}}\ , \nonumber
\eeqn
where the conformal gauge has been chosen
\be
g_{\mu \nu}^{(2)} = e^\sigma e^{2 \dzzw} \eta_{\mu \nu}\ .
\ee
This is of the form given in \cite{GHR,GGW} and the terms proportional 
to $\epsilon^{\mu \nu}$ display the fact that the $\Phi^\Sigma$ 
reside in chiral multiplets whereas the $t^A$ are members of 
twisted chiral multiplets. These are the coordinates which are relevant for this paper.

It is however possible to express the $D=2$ effective action 
(\ref{2a2p}) by chiral multiplets only. For this purpose it is
 necessary to dualise the scalars $a^A$. This is 
possible, because they only appear via their 
'field strength' $\pu a^A$ in (\ref{2a2p}).  
One adds  
$-F^A_\mu \po P^A$ 
to (\ref{2a2p}), where $P^A$ is a Lagrange multiplier and 
$F^{A \rho} \equiv  \epsilon^{\rho \mu} \pu a^A $. Then one
eliminates $F^A$ via its equation of motion
in favour of $P^A$.
Rescaling the $M^A$ according to\footnote{
The  $\tm^A$  are precisely the K\"ahler moduli 
of M-theory used in \cite{HL}.} 
\be
\tm^A = e^{-2/3 \dzze} M^A 
\label{tm}
\ee
rescales the couplings as follows
\beqn
\tg_{AB} & = & G_{AB} e^{4/3 \dzze}\  , \non
\tg_{I \bar J} & = & G_{I \bar J} e^{-2/3 \dzze}\  , \\
\tvol & = & e^{-8/3 \dzze} \vol =  e^{-2/3 \dzze} 
e^{-2 \dzzw}\ . \nonumber
\eeqn
Finally, we choose a different conformal gauge
for the two-dimensional metric
\be
g^{(2)}_{\mu \nu} = e^{2 \dzze} e^{4 \dzzw} e^\sigma 
\eta_{\mu \nu}\ .
\ee
Inserting these field redefinitions into (\ref{2a2p}) 
and performing the duality transformation yields
\beqn
{\cal L}_{IIA}^{(2)} & = & e^{-2 \dzzw} 
\left[ -\frac{1}{2} \pu \po \sigma 
- G_{\alpha \bar{\beta}} \pu Z^\alpha 
\po \bar{Z}^{\bar{\beta}} 
- \tvol^{-1} \tg_{I \bar{J}} D_\mu N^I 
D^\mu \bar{N}^{\bar{J}} \right. \non
& & \mbox{} - \frac{1}{2} \pu \ln \tvol 
\po \ln \tvol - \frac{1}{2} \tg_{AB} 
\pu \tm^A \po \tm^B 
\label{IIA2d} \\
& & \mbox{} -\frac{1}{2 \tvol^2} 
\Big(\pu P^{A} + \frac{1}{4} d_{A K \bar{L}} 
\left( N^{K} D_\mu \bar{N}^{\bar{L}} - 
D_\mu N^K \bar{N}^{\bar{L}} \right) \Big) \non
& & \left. \tg^{-1}_{AB} \Big(\po P^{B} + \frac{1}{4} 
d_{B I \bar{J}} 
\left( N^{I} D^\mu \bar{N}^{\bar{J}} - 
D^\mu N^I \bar{N}^{\bar{J}} \right) \Big) \right].
\nonumber
\eeqn
This is 
exactly the effective action one gets by 
first reducing $D=11$ supergravity on the Calabi-Yau 
fourfold to $D=3$ as performed in \cite{HL} and then 
reducing on a further circle. To see this one has to
 use the relation between the $D=2$ type IIA dilaton 
and the radius $r$ of the circle measured in the $D=3$ 
Einstein-frame metric $r = e^{-2 \dzzw}$. In addition 
one has to choose the $D=2$ M-theory metric as
\be
g^{(2)}_{{\rm M} \mu \nu} = e^\sigma \eta_{\mu \nu}\ .
\ee 
It is clear from the discussion in \cite{HL} that one can 
write (\ref{IIA2d}) in the form
\be\label{l22a}
{\cal L}_{IIA}^{(2)} = e^{-2 \dzzw} 
\left[ -\frac{1}{2} \pu \po \sigma - G_{\bar \Lambda 
\Sigma} \pu \bar Z^{\bar \Lambda} \po Z^\Sigma \right]\ ,
\ee
where $G_{\bar\Lambda\Sigma} =
\bar\partial_{\bar\Lambda}\partial_{\Sigma}
K^{(3)}_{\rm M}$ and $Z^\Sigma = 
\{ T^{A}, \nN^I, Z^\alpha \}$ with 
\beqn
K^{(3)}_{\rm M} & = & -\ln[\int_{Y_4}\Omega\wedge\bar \Omega]
 - \ln\Big[\Xi^A \vol G^{-1}_{AB} \Xi^B\Big]\ ,\non
\label{k3m}
\Xi^A & \equiv &
 -i \Big( T^{A} - \bar{T}^{A} -
\frac{1}{2 \sqrt{8}}  d_{A M \bar{L}} 
\Gi^{-1}_{\bar{J} M}
\Gi^{-1}_{\bar{L} I}
\nN^{I} \bar{\nN} \, \! ^{\bar{J}} 
- \frac{i}{\sqrt{8}} (\omega_{A I K} \nN^I \nN^K 
+\bar{\omega}_{A \bar{J} \bar{L}} 
\bar{\nN} \, \! ^{\bar{J}} 
\bar{\nN} \, \! ^{\bar{L}}) \Big)\ , \non
\label{Xi} 
T^{A} &=& \frac{1}{\sqrt{8}} 
\Big( - P^{A} + i \tvol \tg_{AB} \tm^{B} 
+\frac{1}{4} d_{A M \bar{L}} \Gi^{-1}_{\bar{J} M}
\Gi^{-1}_{\bar{L} I}
\nN^{I} \bar{\nN} \, \! ^{\bar{J}} 
+ i \omega_{A I K} \nN^I \nN^K \Big)\ . \label{TA} 
\eeqn 
In the coordinates
$Z^\Sigma$ the moduli space in $D=2$ is 
K\"ahler and has the same K\"ahler potential as 
in $D=3$.\footnote{Note that in comparison with \cite{HL} 
we have here redefined the coordinates $T^A \rightarrow i T^A$.}  
Therefore the discussion of duality between the 
heterotic and type IIA theory in $D=2$ using the variables of (\ref{l23}) and 
(\ref{l22a}) proceeds exactly along the same lines as in \cite{HL}. 
Finally comparing (\ref{l22a}) with (\ref{l23}) shows 
\be \label{deqd}
e^{-2 \dhz} = e^{-2 \dzzw}\ .
\ee
This 
can also heuristically be derived
from the duality between the heterotic and type IIA theory  
in $D=6$. This duality maps the string-frame metrics 
according to $g^{(6)}_{\rm het} = e^{2 \Phi_{\rm het}^{(6)}} g_{\rm IIA}^{(6)}$
and the dilatons are related via 
$\Phi_{\rm het}^{(6)} = -\Phi_{\rm IIA}^{(6)}$. 
Starting from the effective action in $D=6$ and 
compactifying further on a four-dimensional manifold $B$ one
derives 
\beqn
S & = & \int d^6x \sqrt{-g^{(6)}_{\rm het}} e^{-2 \Phi_{\rm het}^{(6)}} 
(R_{\rm het}^{(6)} + \ldots) \non
& = & \int d^2x \sqrt{-g^{(2)}_{\rm het}} e^{-2 \Phi_{\rm het}^{(6)}} \vol_B^{\rm
  het}
(R_{\rm het}^{(2)} + \ldots) \non
& = & \int d^2x \sqrt{-g^{(2)}_{\rm het}} e^{-2 \dhz}
(R_{\rm het}^{(2)} + \ldots)\ . 
\eeqn
On the other hand the duality relations given above
yield $e^{-2 \Phi_{\rm het}^{(6)}} \vol_B^{\rm het} = e^{2 \Phi_{\rm het}^{(6)}} 
\vol_B^{\rm IIA} =
e^{-2 \Phi_{\rm IIA}^{(6)}} \vol_B^{\rm IIA} = e^{-2 \dzzw}$, where 
all volumes are measured in the respective string-frame metrics.

\newpage

\end{document}